\documentclass[pre,twocolumn,showpacs,preprintnumbers,amsmath,amssymb,floatfix]{revtex4}
\usepackage{graphicx}
\usepackage{dcolumn}
\usepackage{bm}
\usepackage{xspace}
\newcommand{\eq}[1]{Eq.~(\ref{#1})}
\newcommand{\taus}{\ensuremath{\tau_\text{S}}\xspace}
\newcommand{\taum}{\ensuremath{\tau_\text{M}}\xspace}
\newcommand{\tauf}{\ensuremath{\tau_\text{flip}}\xspace}
\newcommand{\dd}{\ensuremath{\mathrm{d}}\xspace}
\newcommand{\M}{\ensuremath{\mathcal M}\xspace}
\newcommand{\avstress}{\ensuremath{\langle\sigma\rangle}\xspace}
\newcommand{\beq}{\begin{equation}}
\newcommand{\eeq}{\end{equation}}
\begin{document}
%
%
\title{A minimal model for chaotic shear banding in shear-thickening fluids}
\author{A.\ Aradian}
\altaffiliation[Also at ]{Centre de Recherche Paul Pascal, CNRS,
Avenue A.~Schweitzer, 33600 Pessac, France}
\email{aradian@crpp-bordeaux.cnrs.fr}
\author{M.~E.\ Cates}%
\affiliation{SUPA, School of Physics, University of Edinburgh, JCMB
Kings Buildings, Edinburgh EH9 3JZ, UK}%
\date{\today}
\begin{abstract}
We present a minimal model for spatiotemporal oscillation and
rheochaos in shear-thickening complex fluids at zero Reynolds
number. In the model, a tendency towards inhomogeneous flows in
the form of shear bands combines with a slow structural dynamics,
modelled by delayed stress relaxation. Using Fourier-space
numerics, we study the nonequilibrium `phase diagram' of the fluid
as a function of a steady mean (spatially averaged) stress, and of
the relaxation time for structural relaxation. We find several
distinct regions of periodic behavior (oscillating bands,
travelling bands, and more complex oscillations) and also regions
of spatiotemporal rheochaos. A low-dimensional truncation of the
model retains the important physical features of the full model
(including rheochaos) despite the suppression of sharply defined
interfaces between shear bands. Our model maps onto the
FitzHugh-Nagumo model for neural network dynamics, with an unusual
form of long-range coupling.
\end{abstract}
\maketitle
\section{\label{Introduction}Introduction}
Complex fluids have long been known to show strong coupling
between structure and flow. This leads to viscoelasticity in both
the linear and the nonlinear response. The latter can include both
shear thinning and shear thickening, with upward and downward
curvature respectively in the steady state flow curve
$\sigma(\dot\gamma)$ of shear stress against strain rate. Polymers
are usually shear thinning, whereas viscoelastic micellar systems
can be either thinning or thickening, as can dense colloidal
suspensions \cite{Larson,CatesHouches}. In extreme cases of shear
thinning, where $\sigma(\dot\gamma)$ becomes nonmonotonic, steady
flow is mechanically unstable on the decreasing portion of the
curve. Stability can sometimes be restored by shear-banding, in
which bands of material with unequal strain rate $\dot\gamma$ but
equal stress $\sigma$ coexist, with interface normals in the shear
gradient direction \cite{Spenley,LuOlmstedBall}. For extreme shear
thickening, the same applies interchanging $\sigma$ and
$\dot\gamma$, with band interface normals now in the vorticity
direction \cite{OlmstedEPL,OlmstedGoveas}.

It has recently become clear, however, that in some complex fluids
(as outlined below) parameter regimes exist where the constitutive
response to steady driving is intrinsically unsteady. This entails
a different (or at least stronger) dynamical instability to the
one present in shear-banding. These instabilities are not
transient phenomena: they rather indicate the presence of
complicated dynamical states within the flow diagrams of complex
fluids. In principle there can be many sources of dynamical
instability: under appropriate conditions, complex fluids---like
simple fluids---are subject to classical inertial (Taylor-Couette,
turbulence) or thermoconvective (Rayleigh-B\'enard)
instabilities~\cite{DrazinHydro,MannevilleLivre2}. More specific
to complex fluids are stick-slip phenomena (like the `spurt' or
`melt fracture' effect~\cite{Larson}) and various elastic
instabilities (so-called `elastic
turbulence'~\cite{ElasticTurbulence}).

Here we address a distinct class of instabilities in which not
only the mechanical response, but also the internal structural
parameters of the fluid, vary in time. Experimentally, such
structural instabilities arise in a variety of systems. Typical
observations fall into two broad types.

The first type of unstable temporal behavior comprises sustained,
periodic oscillations of the shear rate at constant imposed shear
stress, or vice-versa: this has been observed for surfactant
solutions either in worm-like micellar
phases~\cite{Hu,Wheeler,Fischer1,Fischer2,Fischer3} or lamellar
phases~\cite{Wunenburger,Salmon1,Manneville,Courbin}, as well as
in polymer solutions~\cite{Vlassopoulos} and concentrated
colloids~\cite{Laun}. The second type of unsteady behavior is more
complex, with erratic temporal responses of either the shear rate
or shear stress. Such irregular signals have been observed in
worm-like
micelles~\cite{Bandyopadhyay1,Bandyopadhyay2,Callaghan1,Callaghan2},
lamellar (onion) phases~\cite{Manneville,Salmon1,Salmon2} and
concentrated colloids~\cite{Lootens}. In many of these systems,
such erratic responses occur for parameter values enclosed within
those where oscillations arise.

There are strong
indications~\cite{Bandyopadhyay1,Bandyopadhyay2,Salmon1} that
these erratic responses result from a deterministic chaotic
dynamics. A remarkable aspect of chaos in these flows is that the
Reynolds number is virtually zero: the inertial term ${\bm
v}.\nabla {\bm v}$ in the Navier-Stokes equation is negligible.
The instabilities thus stem from constitutive nonlinearity in the
rheology of the fluid, unlike the convective instability that
gives rise to turbulence in Newtonian flows. The term `rheological
chaos', or simply `rheochaos' has been coined for such
behavior~\cite{CatesHead}.

The involvement of the fluid microstructure in these instabilities
was established by monitoring structural observables and showing
that these evolve in concert with the time-dependent rheological
signal. The methods used include birefringence
imaging~\cite{Wheeler,Vlassopoulos}, light
scattering~\cite{Wunenburger,Salmon1} and spatially resolved
NMR~\cite{Callaghan2}. Many (but not quite all) instances of
structural instability occur in shear-stress or shear-rate ranges
close to non-equilibrium transitions between distinct phases or
textures in the fluid, for example the transition from isotropic
to flow-aligned-nematic structures in worm-like
micelles~\cite{Callaghan2}, or the disordered-to-layered packing
transition in multilamellar
onions~\cite{Wunenburger,Salmon1,Manneville,Salmon2}. (There may
also be underlying transitions from a flowing to a jammed state in
colloids~\cite{Lootens}, and from isotropic to string-like
structures in polymer solutions~\cite{Vlassopoulos}.) Structural
instabilities arguably arise when the fluid under flow hesitates
between possible alternative structures near such transitions.

A crucial question is whether structural instabilities are purely
temporal or spatiotemporal in character. Do all points in the
fluid follow the same time evolution, or do different parts of the
fluid have different mechanical and structural states at the same
instant of time? Early experiments on worm-like
micelles~\cite{Wheeler,Fischer1} and polymers~\cite{Vlassopoulos}
pointed towards heterogeneity: optical observations showed
alternating turbid and clear bands. More recent advances allow
spatially and temporally resolved measurements of velocity
profiles within a rheometer. Such experiments, on multilamellar
onions~\cite{Salmon2,Manneville} and worm-like
micelles~\cite{Callaghan2}, unambiguously demonstrate both the
presence and the nature of the heterogeneities: they are
\emph{fluctuating shear bands}. Theoretical models that address
only temporal instability in a spatially uniform system, such as
that of Ref.~\onlinecite{CatesHead}, are therefore of limited
relevance to the experimental situation.

In this paper we extend the model of Ref.~\onlinecite{CatesHead}
to allow for spatial heterogeneity, exploring the resulting
scenario of why shear bands destabilise in such fluids and how
this produces oscillatory and chaotic flows. (Note that we work in
a parameter regime where the flow curve is monotonic, so that
shear bands cannot exist as a time-independent steady state.)

We thereby create a minimal model of spatiotemporal instabilities
in shear-thickening fluids. The model includes an intrinsic
short-time tendency to form shear-bands coupled to a slow
relaxational component dynamics for the fluid microstructure
(modelled as a retarded stress response term). In the purely
temporal model of Ref.~\onlinecite{CatesHead}, oscillations but
not chaos were found (unless a physically unconvincing `double
memory' term was used). By allowing for full spatiotemporal
dynamics, we show that the interplay between the above two factors
gives rise to  several distinct periodic regimes (including
oscillating shear bands and travelling bands) and also regimes of
spatiotemporal rheochaos.  Preliminary accounts of our work were
given in Refs.~\onlinecite{AradianCatesEPL,AradianProceedings}.

Our motivation for studying the case of a fluid that shows
shear-thickening (in itself a widely observed but poorly
understood phenomenon~\cite{HolmesJam,Rings}) is that several of
the above-cited experiments concern such fluids
(Refs.~\cite{Hu,Wheeler,Fischer1,Fischer2,Bandyopadhyay2} for
wormlike micelles; \cite{Vlassopoulos} for polymer solutions;
and~\cite{Laun,Lootens} for colloidal suspensions).

The alternative case of shear-thinning fluids has been recently
addressed by Fielding and Olmsted~\cite{Fielding}, and in less
detail also by ourselves~\cite{AradianProceedings}. The authors of
Ref.~\onlinecite{Fielding} have demonstrated the presence of a
rich dynamics, including rheochaos, in their model. Also closely
related is the work by Chakrabarti et al.~\cite{Ramaswamy} on
nematic liquid crystals which also shows regimes of spatiotemporal
chaos. In what follows, we shall note similarities and differences
between our own work and these other studies.

The rest of this paper is organized as follows. In
section~\ref{model}, we define the model, discuss its physical
assumptions, and relate it to other models (both in rheochaos and
in other fields of study). In section~\ref{qualitative}, we give a
qualitative description of how the model works.
Section~\ref{numerics} explains how the model is solved
numerically. In section~\ref{hightruncation}, we present our
results in the form of a nonequilibrium `phase diagram' and
comment on the various flow regimes encountered.
Section~\ref{lowtruncation} studies a low-dimensional version of
the model which offers further insights into the nature of
rheochaos. In Section~\ref{concsec} we summarize our findings and
their generic implications for the physics of rheological
instabilities in complex fluids.
\section{\label{model}The model}
Our model has two main physical ingredients, as follows:

\emph{(i)}~the fluid has a tendency to form  shear bands;

\emph{(ii)}~there are slow structural modes in the fluid whose
delayed relaxation modifies the evolution of the stress.

The latter ingredient is central to our description. When some
structural mode is disturbed by the flow, it will relax on a
timescale that is distinct from the stress relaxation time in the
system. Below, for simplicity, we will consider a single timescale
\taus for structural relaxation, with stress relaxation
assimilated into the usual Maxwell time \taum. We focus on the
case where structural modes relax at least as slowly as the stress
itself would do at fixed structure ($\taus > \taum$).

Because of the disordered energy landscape and metastability
intrinsic to many complex fluids~\cite{StAndrews,SGRPRL,HeadSGR},
such slow dynamics is commonplace. Candidates for slow structural
modes include the mean length or the local gel fraction in
worm-like micelles; local composition variables (e.g., colloidal
volume fraction); and `fluidity' parameters reflecting, e.g., a
local bonding state~\cite{Fluidity,AradianProceedings}. An
involvement of slowly evolving structure has been shown in many
experimental
cases~\cite{Wunenburger,Salmon1,Manneville,Courbin,Vlassopoulos,
Hu,Wheeler,Fischer1,Fischer2,Fischer3}.
\subsection{\label{equation}Model equations}
\begin{figure}
\includegraphics[width=.44\textwidth,clip=true]{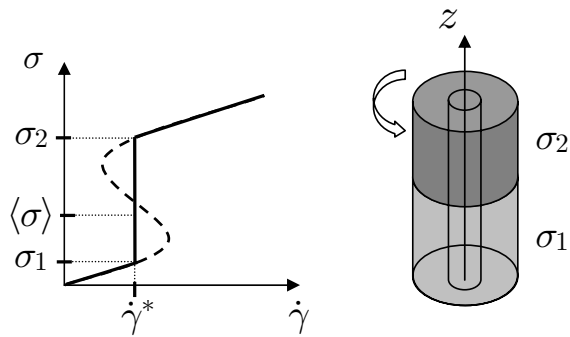}
\caption{\label{VorticityBands} Vorticity shear bands in a Couette
cell~\cite{OlmstedEPL,OlmstedGoveas}. These occur in
shear-thickening materials with an S-shaped flow curve ($\sigma$
vs.\ $\dot\gamma$). A given mechanical torque on the Couette
imposes a mean (spatially averaged) stress \avstress. Any
\avstress within the unstable portion of the flow curve is
unstable toward shear bands (with different local microstructures)
along the vorticity direction $z$, depicted here as a clear and a
turbid band. These coexist at a common shear rate
$\dot\gamma_\text{c}$ (whose value is fixed by gradient
terms~\cite{LuOlmstedBall}) but at different stresses $\sigma_1$
and $\sigma_2$. The amount of each band is such that the weighted
mean of $\sigma_1$ and $\sigma_2$ matches the externally imposed
mean \avstress.}
\end{figure}
We consider the situation of a fluid under pure shear, and we
assume, as usual~\cite{CatesHouches}, that the shear stress
$\sigma$ decouples from other stress components, and depends only
on the rate of shear strain $\dot\gamma$. We will restrict our
study to one-dimensional spatial heterogeneity in the fluid, in
the vorticity direction (perpendicular both to the velocity and
the velocity gradient); in cylindrical Couette geometry, which we
will always refer to in the following, this corresponds to the
axial coordinate denoted $z$. (For 3D-related effects in
situations of shear-banding, see~\cite{FieldingPlanarInterface}.)

This choice for the heterogeneity direction is motivated by the
classical geometry of steady shear bands in thickening
materials~\cite{OlmstedEPL,OlmstedGoveas} where, as shown in
Fig.~\ref{VorticityBands}, bands are stacked one onto another in
the vorticity direction (vorticity shear bands). Note that, under
these assumptions, the shear flow is homogeneous within each slice
of height $z$, as required by the low-Reynolds limit. (We neglect
any small variation in the velocity gradient direction caused by
the curvature of the cell. Without fluid inertia, the stress
cannot vary in this direction.)

In all of the following, we shall work under conditions of imposed
torque, i.e., under an imposed value of the mean (spatial average)
of the stress \avstress: this is the usual situation for vorticity
shear bands, shown in Fig.~\ref{VorticityBands}. [We also studied
the behaviour of our model under conditions of imposed shear rate
(results not shown) but no spatiotemporal effect was observed, see
note~\footnote{For completeness, we also explored the behaviour of
the model of \eq{spaceCHA} under conditions (unusual for vorticity
banding) of imposed shear rate. In all runs, without exception,
the response proved purely temporal: in contrast with our results
under imposed stress discussed in the main text, here the dynamics
in the system converges very early on to a situation where all
points are in the same state at any given instant of time and
oscillate in synchrony over space. The only observable instability
thus corresponds to global, periodic oscillations of the system
(no shear bands nor rheochaos). The effective dynamics for each
point in the cell is simply that seen in the earlier, temporal
version of the model proposed in Ref.~\onlinecite{CatesHead}.}.]

In our model, the shear stress $\sigma(z,t)$ at time $t$ evolves
according to the following equation of motion:
\begin{equation}
\label{spaceCHA} \dot\sigma(z,t)= \dot\gamma(t) -R(\sigma) - \lambda
\int_{-\infty}^t\hspace{-.3cm} \M(t-t') \,\sigma(z,t')\,
\mathrm{d}t' +\kappa\nabla^2\sigma
\end{equation}
where
\begin{equation}
\label{defRandM} R(\sigma)=a\sigma-b\sigma^2+c\sigma^3
\quad\text{and}\quad\M(u)=\frac{1}{\taus}\,e^{-u/\taus}
\end{equation}
Eqs.~(\ref{spaceCHA})--(\ref{defRandM}) are non-dimensional: the
transient elastic modulus is taken as the unit for stress, and
$H$, the axial extent of the Couette cell (vertical in
Fig.~\ref{VorticityBands}), the unit of length. Several choices
are possible for the time unit; the simplest one is the fluid's
Maxwell time $\taum$. However, as we will always be dealing with
times longer than \taum, we choose instead $100\,\taum$ as a time
unit. [This amounts to setting $\taum=0.01$ and, via \eq{defTaum},
$a=100$ throughout what follows; we do this.] The reader is
referred to Appendix~\ref{NonDimensional} for details on the
non-dimensionalisation procedure.

Note that $\dot\sigma=\partial\sigma/\partial t$ in
\eq{spaceCHA} is a local time derivative. We also emphasise that
the shear rate $\dot\gamma$ in \eq{spaceCHA} is \emph{uniform}:
the moving wall of the rotor imposes the same velocity for all
heights $z$, and $\dot\gamma(z,t)=\dot\gamma(t)$ only.

\begin{figure}
\includegraphics[width=.44\textwidth]{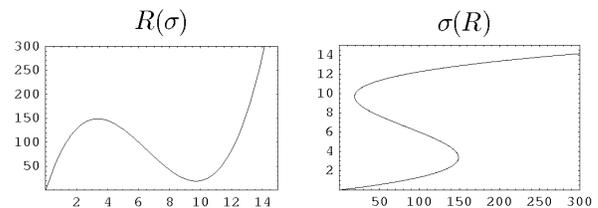}
\caption{\label{RandSigma} Plots of $R(\sigma)$ and the inverse
function $\sigma(R)$. The S-shape of the latter encodes the
tendency of the fluid to form vorticity shear bands.}
\end{figure}

The terms in Eqs.~(\ref{spaceCHA}) and (\ref{defRandM}) have the
following significance. $R(\sigma)$ describes instantaneous,
nonlinear stress relaxation. As defined by
\eq{defRandM}, $R$ is a third-order polynomial, with the positive
constants $a$, $b$ and $c$ chosen so that $R(\sigma)>0$ for
$\sigma>0$, and so that the inverse function, $\sigma(R)$, is an
S-shape (see Fig.~\ref{RandSigma}). We choose $a=100$, $b=20$,
$c=1.02$ in this paper; setting $c = 1$ would give a physically
inappropriate zero of $R(\sigma)$ at $\sigma=10$. The $R(\sigma)$
term encodes ingredient \emph{(i)} of the model, i.e., the
tendency of the fluid to form vorticity shear-bands. This tendency
is instantaneous, but frustrated by structural relaxation effects
--- see Fig.~\ref{FlowCurves}. Note that, with our choice of
units, $a$ controls linear stress relaxation in the fluid, and
hence determines the Maxwell time \taum:
\begin{equation}
\label{defTaum}
\taum=1/a
\end{equation}
Setting $b=c=\lambda=\kappa =0$ in \eq{spaceCHA} indeed recovers
the classical Maxwell model for linear viscoelasticity.

The integral term in \eq{spaceCHA} corresponds to our second
physical ingredient, and represents retarded stress relaxation due
to slow structural reorganisation in the fluid; $\lambda$ is a
positive constant governing the strength of this term. For the
sake of simplicity, retarded relaxation is chosen linear in past
stresses; but note that, as discussed in
Ref.~\onlinecite{CatesHead}, a similar form can be obtained by
introducing an explicitly structure-dependent $R(\sigma)$ and then
linearizing this structure-dependence. The memory kernel
$\mathcal{M}$ is in principle any decaying
function~\cite{CatesHead}. However, in \eq{defRandM}, we choose it
mono-exponential with characteristic time \taus. This choice
permits a much simpler, fully differential representation that we
exploit in our numerics below. As stated previously, we take
structural relaxations to be slow compared to the intrinsic time
scale of stress relaxation; we study mainly $4<{\taus}/{\taum}<
10^4$. (This contrasts somewhat with the model of
Ref.~\onlinecite{Fielding} where ${\taus}/{\taum}\simeq 1$.)

Finally, the last term of \eq{spaceCHA} assigns to stress a
diffusivity $\kappa$. By analogy with the classical Cahn-Hilliard
model for phase separation~\cite{ChaikinLubensky}, this can be
viewed as representing the cost of maintaining interfaces in
inhomogeneous states. Such minimal nonlocality in the constitutive
model is known to be physically crucial. For example, in
steady-state shear bands it drives a selection mechanism, among
all possibilities on a given flow curve like that of
Fig.~\ref{VorticityBands}, for the coexistence strain rate
$\dot\gamma_\text{c}$~\cite{LuOlmstedBall}.

A superficial comparison of Eq.~(\ref{spaceCHA}) with
Ref.~\onlinecite{CatesHead} might suggest that our work is new
only in the addition of this diffusive term. But of course we
simultaneously upgrade the dimensionality of the problem from
$\sigma(t)$ in that work to $\sigma(z,t)$ here; this allows vastly
richer behavior, in better correspondence with the experimental
facts, to emerge.

We now recast Eqs.~(\ref{spaceCHA})--(\ref{defRandM}) into a
purely differential form~\cite{CatesHead}. We define a new
variable
\beq
\label{memory} m(z,t)=\int_{-\infty}^t \M(t-t') \,\sigma(z,t')\,
\mathrm{d}t'
\eeq
which we call the `memory', and which contains the structural part
of the stress relaxation. We can then rewrite Eq.~(\ref{spaceCHA})
as an exactly equivalent differential system
\begin{eqnarray}
\label{RheologicalEquation}
\dot\sigma&=&\dot\gamma-R(\sigma)-\lambda m +
\kappa\nabla^2\sigma\\
\label{StructuralEquation}
\dot m&=&-\frac{m-\sigma}{\taus}
\end{eqnarray}
Equation~(\ref{StructuralEquation}) states that the memory is
always relaxing towards the current and local value of the stress
$\sigma(z,t)$ with a (slow) rate $\taus^{-1}$.

In its differential form, the model can be interpreted as follows.
Equation~(\ref{RheologicalEquation}) is a nonlinear rheological
equation controlling the temporal evolution of the stress, where
one of the participating variables is structural by nature (the
memory $m$); this structural variable is subject to a distinct
dynamics, governed by a `structural equation',
\eq{StructuralEquation}, with a different relaxation time \taus.
The coupling of the two dynamics, mechanical and structural, is
the source of the instability.

\subsection{Relation to other models\label{ComparisonModels}}
%
%

Viewed in this way, our model belongs to a larger class of models
that have been recently proposed to describe the dynamics of
complex fluids.

The models of Refs.~\cite{Fielding} and
\cite{AradianProceedings} for
shear-thinning solutions of worm-like micelles work on a similar
scheme: there, the `structural variable' is the mean chain length
of the micelles~\cite{Fielding}, or simply the Maxwell time of the
fluid itself~\cite{AradianProceedings}. In both cases, its
evolution is governed by a differential `structural equation' akin
to \eq{StructuralEquation}. Belonging to the same family, albeit
coming from a somewhat different perspective, Derec et al.
proposed a model for the fluidisation transition in
pastes~\cite{Fluidity}, where a classical rheological equation is
supplemented by a structural equation describing a `fluidity'
parameter.

From a more formal point of view, all these models are related to
the prey-predator and reaction-diffusion models commonly found in
nonlinear physics and mathematical biology~\cite{Murray}. (In our
case, $\sigma$ and $m$ are the two competing species). This type
of non-linear models are known to yield complex spatiotemporal
behavior~\cite{Murray}, thus it should not come as a surprise that
the related models used in the field of complex fluids do also.

It is interesting to note that our
Eqs.~(\ref{RheologicalEquation})--(\ref{StructuralEquation}) in
fact map exactly onto a well-known model of nonlinear physics, the
FitzHugh-Nagumo
model~\cite{FitzHugh,Nagumo,Murray,KeenerSneyd,Rocsoreanu}. The
simplest version of the latter model was developed to describe
electrical activity in a single neuron, where the axon membrane
potential is a fast variable (also called `excitation variable',
analogous to $\sigma$ in our model), and is coupled to the
dynamics of a slow variable (or `recovery variable', analogous to
$m$), related to the activity of sodium ion channels. This system
is known to produce van-der-Pol-like (purely temporal)
oscillations, like those described for complex fluids in the
earlier, spatially homogenised, version of our
model~\cite{CatesHead}.

In recent years, spatially inhomogeneous extensions of the
FitzHugh-Nagumo have been used to study the collective properties
of interacting networks of neurons, and other related assemblies
of coupled nonlinear oscillators, focussing on the competing
effect of local and global couplings~\cite{FHN1,FHN2,FHN3,FHN4}.
The model of
Eqs.~(\ref{RheologicalEquation})--(\ref{StructuralEquation}) is of
just this type: the local coupling is supplied by the diffusion
term $\kappa\nabla^2\sigma$, and the global coupling by the
constraint that the spatial mean stress \avstress is externally
fixed.

We remark however that:
\emph{(a)}~in our model, the mean value of the stress is
fixed, while in neural networks where a coupling to the mean
membrane potential is implemented~\cite{FHN1,FHN3,FHN4}, its value
fluctuates in response to the collective dynamics in the model;
\emph{(b)}~possibly for this reason, several of the spatiotemporal
patterns reported below (in particular oscillatory shear bands)
have not to our knowledge been reported in the literature on
FitzHugh-Nagumo networks; \emph{(c)}~an active research topic in
neural networks is the effect of noise on global behavior,
including the possibility to trigger chaotic dynamics~\cite{FHN4}.
Such a role for noise (either thermal, or mechanical) is neglected
in most models of rheochaos but would constitute an interesting
avenue for future research.
\section{Qualitative features\label{qualitative}}
We now explain some qualitative features of our model, focussing
on the origin of the dynamical instability, and on a physically
important scaling property. We also briefly discuss how the values
of our model parameters may be estimated from experiment.
\subsection{Flow curves}
We start by computing the flow curves for the fluid, i.e., the
relation between $\sigma$ and $\dot\gamma$ in steady-state flow.

\begin{figure}
\includegraphics[width=.44\textwidth]{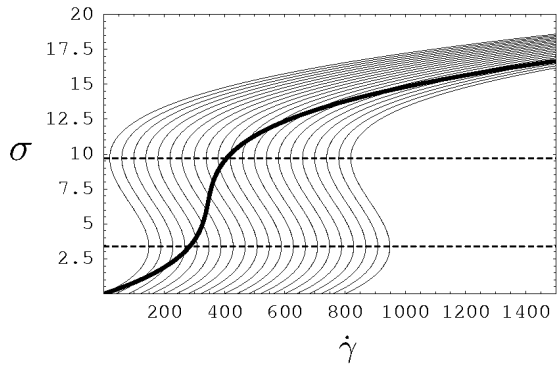}
\caption{\label{FlowCurves} Steady-state flow curve (thick line), and
underlying short-term flow curves (thin lines, with, from left to
right, $m=0,1,2,\ldots,20$). The stress range between the dotted
lines corresponds to the unstable region, here deduced for
$\taus=100$ from \eq{InstabilityCondition}. Numerical parameters:
$\lambda=40$, $a=1/\taum=100$, $b=20$, $c=1.02$, $\kappa=0.01$,
$H=1$.}
\end{figure}

From Eqs.~(\ref{RheologicalEquation})--(\ref{StructuralEquation}),
the flow curve for steady-state, homogeneous flows is given by
setting $\dot\sigma=\dot m=\nabla^2\sigma=0$.
Equation~(\ref{StructuralEquation}) then yields $m=\sigma$: at
each point on the curve, the memory has relaxed to the
steady-state stress. This can be substituted in
\eq{RheologicalEquation}, to provide the equation for the steady-state flow
curve~\cite{CatesHead}:
\beq
\label{LongTermFlowCurve}
\dot\gamma=R(\sigma)+\lambda\sigma 
\eeq
When $R'(\sigma)+\lambda\geq0$ for all values of $\sigma$ (a
condition which we shall always take to hold below) the flow curve
is monotonically increasing, as in the thick curve in
Fig.~\ref{FlowCurves}. In contrast with rheological constitutive
models that do not show structural coupling, this monotonicity
does not ensure mechanical stability of homogeneous
flows~\cite{CatesHead}.

The qualitative reason for instability of the monotonic flow curve
is that the steady states associated to this curve only arise when
the memory $m$ has relaxed, i.e., for time scales beyond \taus.
For times much shorter than \taus, the fluid instead behaves as if
the memory $m$ was frozen. Thus, at short time scales, the fluid
lives on one of a set of `instantaneous flow curves' for which
stress has relaxed, but not structure. These instantaneous curves
have $\dot\sigma=0$ at fixed $m$, giving
\beq
\label{ShortTermFlowCurve}
\dot\gamma=R(\sigma)+\lambda
m\qquad\text{(short term)},
\eeq
where $m$ is a parameter. Such curves for $m=0,\ldots,20$ are
plotted in Fig.~\ref{FlowCurves}, and are not monotonic. As the
memory slowly relaxes, the fluid drifts from one such curve to
another; \eq{LongTermFlowCurve} can be reconstructed by picking,
for each value of the stress $\sigma$, the corresponding point on
the particular instantaneous curve that has $m=\sigma$. But for
our choice of non-monotonic $R$, the fluid has an instantaneous
tendency to form shear-banded flow. This impedes the establishment
of a steady state and instability can arise even when the
steady-state flow curve is monotonic~\cite{CatesHead}.
\subsection{Origins of the dynamical instability\label{OriginsInstability}}

As just discussed, the existence of decreasing portions of the
short-term flow curve, in contradiction with the steady-state flow
curve, is a source of instability. In the homogeneous version of
the model (no $z$-dependence), this causes temporal instability in
the form of van-der-Pol-type oscillations~\cite{CatesHead}.
Essentially, the mechanism for such oscillations is as follows:
starting for instance from a situation of high stress as compared
to the memory ($\sigma>m$), the memory $m$ will start to grow as
dictated by
\eq{StructuralEquation}; this growth of $m$ in turn brings a
decrease in the value of $\sigma$ through
\eq{RheologicalEquation}; adapting to this, $m$
then decreases, which increases $\sigma$, thereby recovering the
initial high-stress situation and starting the cycle anew.

When spatial dependence is added, this basic temporal oscillation
becomes compounded with shear banding. It is easily seen how this
can give complex spatiotemporal dynamics: looking at
Fig.~\ref{FlowCurves}, if one imposes a mean stress \avstress
chosen within the unstable region, the system decomposes into
several shear bands with unequal local stress, as depicted on
Fig.~\ref{VorticityBands}. Unlike the classical situation of
Fig.~\ref{VorticityBands}, here the van-der-Pol-type temporal
oscillation  rules out steady states for the shear bands: a local
unstable dynamics engages. Simultaneously, the bands have to
match, all together, the constraint on \avstress. This creates
long-range couplings between the bands (when one oscillates,
another has to compensate), considerably complicating the
spatiotemporal behavior.
\subsection{Linear stability analysis}
We have performed a standard linear instability calculation on the
model~(\ref{RheologicalEquation})--(\ref{StructuralEquation}),
with the following results:

\emph{(i)}~for an externally imposed value of the mean stress~\avstress,
instability for stress evolution arises only if
\beq
\label{InstabilityCondition} R'(\avstress)+\frac{1}{\taus}<0.
\eeq

\emph{(ii)}~when the above instability condition is obeyed, only
stress modes with wavevectors $q$ in the range
\beq
0\leq q \leq \sqrt{-\frac{R'(\avstress)+1/\taus}{\kappa}}
\eeq
are unstable. Also, modes with smaller wavevectors have a larger
growth rate: when the uniform $q=0$ mode is constrained, the
lowest nonzero $q$-mode grows fastest.

The instability condition~\emph{(i)} is the same as for the purely
temporal version of the model~\cite{CatesHead}; it states that
instability can arise only in regions of decreasing $R$ (or
equivalently, within the decreasing portions of the short-term
flow curves in Fig.~\ref{FlowCurves}), and that it is facilitated
as $\taus$ becomes larger or impeded as it becomes shorter.

The second result~\emph{(ii)} is a natural consequence of stress
diffusion with diffusivity $\kappa$; this kills off fluctuations
with too large $q$ (small wavelengths) and prevents such modes
from becoming unstable.
\subsection{Scaling property of the model}
Let us assume that a stress response $\sigma(z,t)$ and shear rate
$\dot\gamma(t)$ are obtained as the solutions of the model of
Eqs.~(\ref{RheologicalEquation})--(\ref{StructuralEquation}),
under conditions of imposed mean stress $\avstress$ and with a
given set of parameters $a=\taum^{\;-1}$, $b$, $c$, $\lambda$,
$\taus$, $\kappa$. Then, choosing any scaling factor $\alpha$, the
scaled responses $\alpha\sigma(z,t)$ and $\alpha\dot\gamma(t)$
would also be obtained as the solutions of the model if one
applied a mean stress $\alpha\avstress$ and used a scaled set of
parameters $a=\taum^{\;-1}$, $b/\alpha$, $c/\alpha^2$, $\lambda$,
$\taus$, $\kappa$.

The proof of this scaling property is elementary and left to the
reader. It means that, although all numerical results given below
are found with one specific set of model parameters ($\lambda=40$,
$a=1/\taum=100$, $b=20$, $c=1.02$, $\kappa=0.01$, $H=1$), they in
fact describe the behavior of a one-parameter manifold of
parameters governed by the scale transformation parameter
$\alpha$.

Secondly, for our chosen parameter set, the observed stresses in
both oscillatory and chaotic regimes are very large (of order 10
in dimensionless units, i.e., ten times the elastic modulus of the
fluid). While such large amplitudes of the elastic strain may be
relevant for certain types of fluids, no direct importance should
be attached to this: one can scale down numerical values of the
stress to around, say, unity, keeping exactly the same oscillatory
or chaotic dynamics. Note however that this will not change the
ratio of stress fluctuation to mean stress; this is always large
for our parameter settings. (The parameter space is vast, and we
have not explored values for which this ratio might be reduced.)
\subsection{Experimental estimates for model parameters\label{ExperimentalValues}}
Our model involves a set of phenomenological model parameters
whose numerical values are a priori unknown. One way around this
difficulty is try to extract estimates for these values using
appropriate experiments: it is thus useful, at this point, to
sketch some of the possible strategies---keeping in mind, however,
that our minimal model is intended essentially for a qualitative
exploration of the physics at work in complex fluids not for an
effective fitting tool for specific nonlinear materials.

(It is also recalled that numerical values used in the article are
non-dimensional; see Appendix~\ref{NonDimensional} for the
procedure to convert dimensional quantities, as obtained from
experiments, to non-dimensional ones.)

A first experimental strategy is as follows: starting from a state
of complete rest (where the structural memory has fully relaxed,
$m=0$), and assuming that the structural time $\taus$ is slow
enough, one may be able to measure the instantaneous flow curve at
zero memory [\eq{ShortTermFlowCurve}] using a moderately rapid
ramp up of the applied stress (e.g., sample the whole curve within
a few minutes). Of course, the obtained curve would be unstable
(fig.~\ref{FlowCurves}), showing a conventional `discontinuous
shear-thickening' scenario with a shear-rate plateau and classical
(steady-state) shear bands; but the information collected on the
plateau (critical shear rate, lower and upper values of the stress
at the jump, \ldots) would probably be sufficient to estimate the
short-term parameters $a$, $b$, $c$, and thereby determine the
function $R(\sigma)$ in the model.

Following this determination of $R$, the value of the coupling
parameter $\lambda$ could then be found through a simple
steady-state experiment where a constant stress $\sigma_\text{ss}$
is imposed for a long period a time (in a stable region of the
phase diagram): as stated by \eq{LongTermFlowCurve}, the shear
rate then corresponds to the (known) instantaneous contribution
$R(\sigma_\text{ss})$ plus the delayed contribution $\lambda
\sigma_\text{ss}$. Thus, a measure of
the shear rate will determine the value of $\lambda$, the only
unknown in the equation.

Next, estimating the value of the structural relaxation time
$\taus$ could be done through a relaxation experiment: from a
situation of steady state under a given applied stress, the value
of the stress is suddenly changed to a new constant value (or even
cancelled); if $\taus$ is long compared to other times in the
system, it should be controlling the global relaxation of the
response to the new steady state. Another possibility which could
be explored could be to look at the linear viscosity after a
period of strong shear and/or after a temperature
jump~\cite{Rings}.

The last parameter that needs to be determined is the stress
diffusivity $\kappa$, which in classical, steady-state shear
banding is related to the width of the interfaces between bands.
We expect $\kappa$ to be extremely
small~\cite{FieldingConcentration} (and in any case, much below
the artificially large value of $10^{-2}$ used throughout this
work for computing reasons), although we are not aware of any
experiments allowing to measure it. This should not be too much of
an issue, as our results show that (below a certain threshold) the
behaviour of the model does not qualitatively depend on the actual
value of $\kappa$.
\section{Numerical methods\label{numerics}}
In this section, we detail the numerics techniques used to obtain
the results presented in the rest of the article.
\subsection{Spectral scheme; boundary conditions}
Our numerical scheme expands the stress in Fourier modes, then is
solving for the evolution of these---rather than directly solving
the model equation on a grid of points in direct
space~\cite{Fielding,Ramaswamy}. Such a spectral
scheme~\cite{Boyd} had two main advantages, as follows:
\emph{(a)}~in Fourier space, working under conditions of fixed
mean stress simply corresponds to fixing the value of the uniform,
zeroth mode in the Fourier expansion of the stress (this is much
more difficult to implement in direct-space schemes);
\emph{(b)}~Not only can arbitrarily accurate numerical results be
obtained by keeping enough Fourier modes in the scheme
(`high-order truncation'), but also, by keeping only a minimal
number of modes (`low-order truncation'), one can obtain a reduced
description whose analysis is much simplified. Such a truncation
is more likely to capture the physics of the problem than a
reduced real-space scheme with a spatial grid of only a few
points. We shall present both high-order and low-order numerics in
what follows.

We must also choose appropriate boundary conditions on the system.
We demand zero stress flux at both ends of the Couette cell:
\beq
\label{BoundaryConditions}
\nabla\sigma\rvert_{z=0}=\nabla\sigma\rvert_{z=H}=0
\eeq
To motivate this, note that stress flux arises through the
diffusion term in~\eq{RheologicalEquation}, which is generally
ascribed to material displacement of stress-carrying
elements~\cite{ElKareh}. Our boundary conditions then reflect the
fact that no fluid element leaves through the top nor bottom of
the cell.

Now recall that, for reasons related to the physics of vorticity
shear banding (Fig.~\ref{VorticityBands}), we have the constraint:
\beq
\label{AverageStress}
\avstress=\frac{1}{H}\int_0^H \sigma(z,t)\,\dd z = \text{const.}
\eeq

We now decompose the stress field onto a Fourier basis of spatial
modes compatible with the above boundary conditions (cosines
only):
\beq
\label{StressFourier}
\sigma(z,t)=\sum_{k=0}^{N-1}\sigma_k(t)\cos(k\pi z/H)
\eeq
and note that $\sigma_0(t) = \avstress$ (given that $H = 1$ in our
units). Similarly for the memory term:
\beq
\label{MemoryFourier} m(z,t)=\sum_{k=0}^{N-1} m_k(t)\cos(k\pi
z/H)
\eeq
The series in Eqs.~(\ref{StressFourier}) and~(\ref{MemoryFourier})
only contain the first $N$ modes $\sigma_k(t)$ and $m_k(t)$ of an
infinite Fourier series: the level $N$ of this truncation (known
as a Galerkin truncation~\cite{Boyd}) determine the overall
accuracy of the scheme. (For a given accuracy, the number of modes
that must be kept is often much lower than the number of grid
points in any real-space discretisation~\cite{Boyd}.)

We now project Eqs.~(\ref{StressFourier})--(\ref{MemoryFourier})
onto the Fourier basis to obtain a set of evolution equation for
each of the mode amplitudes $\sigma_k(t)$ and $m_k(t)$. Thanks to
the simple functional form of the model, it is possible to obtain
analytical expressions for the mode equations, which are given in
full in Appendix~\ref{FullAnalytical}. These mode equations are of
the form ($1\leq k
\leq N-1$)
\begin{eqnarray}
\label{ShortModeStressEq}
\dot\sigma_k&=&\text{(coupling terms)} - \lambda m_k - \kappa\,
q_k^2\, \sigma_k\\
\label{ShortModeMemoryEq}
\dot m_k &=&-\frac{m_k-\sigma_k}{\taus}
\end{eqnarray}
where $q_k=k\pi/H$ is the wavevector, and the coupling terms,
stemming from the non-linearity of the instantaneous term
$R(\sigma)$, link the evolution of each $\sigma_k$ mode to all
others. As expected, stress diffusivity damps higher modes via a
linear $\kappa q^2$ term.

Finally, the equations for the $k=0$ uniform modes
$\sigma_0(t)=\avstress$ and $m_0(t)$ are
\begin{eqnarray}
\label{UniformStressEq}
\dot\sigma_0&=&0=\dot\gamma(t)-\frac{1}{H}\int_0^H R\bigl(\sigma(z,t)\bigr)\,\dd z
-\lambda m_0(t)\qquad \\
\label{UniformMemoryEq}
\dot m_0&=&-\frac{\avstress-m_0(t)}{\taus}
\end{eqnarray}
Integrating \eq{UniformMemoryEq} gives [with $m(t=0) =0$, see
below]
\beq
m_0(t)=\avstress (1-e^{-t/\taus})
\eeq
Plugging this expression into \eq{UniformStressEq}, we obtain an
important expression which provides us with the instantaneous
value of the shear rate $\dot\gamma$, once the value for the
stress has been calculated from the Fourier modes:
\beq
\label{ShearRateEq}
\dot\gamma(t)=\frac{1}{H}\int_0^H R\bigl(\sigma(z, t)\bigr)\,\dd z +
\lambda\avstress(1-e^{-t/\taus})
\eeq
The spatial integral in this expression can in fact be carried out
analytically, not numerically (see Appendix~\ref{FullAnalytical}).

The final outcome of the Fourier-Galerkin truncation scheme is
that the partial differential equations
(\ref{RheologicalEquation})--(\ref{StructuralEquation}) in the
original formulation of the model have become a dynamical system
of finite order, containing $2N$ ordinary differential equations
for the modes $\sigma_k(t)$ and $m_k(t)$ as specified by
Eqs.~(\ref{ShortModeStressEq})--(\ref{ShortModeMemoryEq}), plus
the above equations for the uniform modes. The numerical
integration of the ordinary differential
equations~(\ref{ShortModeStressEq})--(\ref{ShortModeMemoryEq}) was
performed with the help of a commercial solver
package~\cite{Mathematica}, using an adaptive time step. Given the
separation of scales between the typical times \taum and
\taus, the adaptive step was needed, to maintain acceptable
computing times.

Initial conditions at $t=0$ were chosen as follows: memory was set
to zero (i.e., $m_k(0)=0$ for all $k$), and initial values for
each stress mode $\sigma_k(0)$ were picked at random between 0 and
$10^{-4}$. The noise amplitude was found not to be essential to
the qualitative results obtained.
\subsection{High-order and low-order truncation}
The behavior of our fluid model was explored using two types of
Fourier-Galerkin schemes: one where the number of modes kept in
Eqs.~(\ref{StressFourier})--(\ref{MemoryFourier}) is high
($N=40$), and one where it is minimal ($N=3$).

For the high-order truncation (Section~\ref{hightruncation}), we
found that keeping $N=40$ modes was numerically accurate, with
reasonable computing times. One criterion in this choice was that
numerics should be able to fully resolve interfaces between shear
bands (the sharpness of interfaces is essentially controlled by
the stress diffusivity $\kappa$; we set it to $10^{-2}$ unless
otherwise stated). A second criterion is that $N$ should be high
enough that results become $N$-independent: this was indeed the
case, with truncations from $N=25$ up all giving similar outputs.
This applies at the level of phase diagrams, etc., but not to
individual trajectories which (at least in chaotic regions of the
phase diagram) can depend on every detail of the numerics.

We also checked the validity of our high-resolution results, for a
given set of parameters, against variations in either initial
conditions for the stress modes, or in the actual sequence of time
steps followed by the adaptative iterator during numerical
integration. We found a limited dependence on these factors from
one run to another. To take this into account, the `phase diagram'
of Fig.~\ref{PhaseDiagram}, where the general behavior of the
model is summarized, was established on the basis of several
independent runs for each point marked in the figure. Thus the
general features of the phase diagram, as well as all conclusions
drawn on the model's global behavior, are reliable.

In the low-order truncation (Section~\ref{lowtruncation}), on the
other hand, our aim was to study the behavior of the model in its
simplest possible spectral representation. In particular, we are
interested in whether low-dimensional chaos was present and what
route led to it. Since three degrees of freedom are required to
allow for chaos, the lowest compatible truncation in our model
corresponds to $N=3$: excluding uniform modes $\sigma_0=\avstress$
and $m_0$, which are dynamically inactive, this leaves a
four-dimensional system involving $\sigma_1$, $\sigma_2$, $m_1$
and $m_2$. In the following sections, we present the results of
the model, first in the high-order truncation, then in the
low-order.
\section{High-order results\label{hightruncation}}
In this section, we present the results that were obtained by
solving the model of
Eqs.~(\ref{ShortModeStressEq})--(\ref{ShortModeMemoryEq}) in the
high-resolution truncation ($N=40$).
\subsection{Phase diagram}
\begin{figure*}
\includegraphics[width=.6\textwidth]{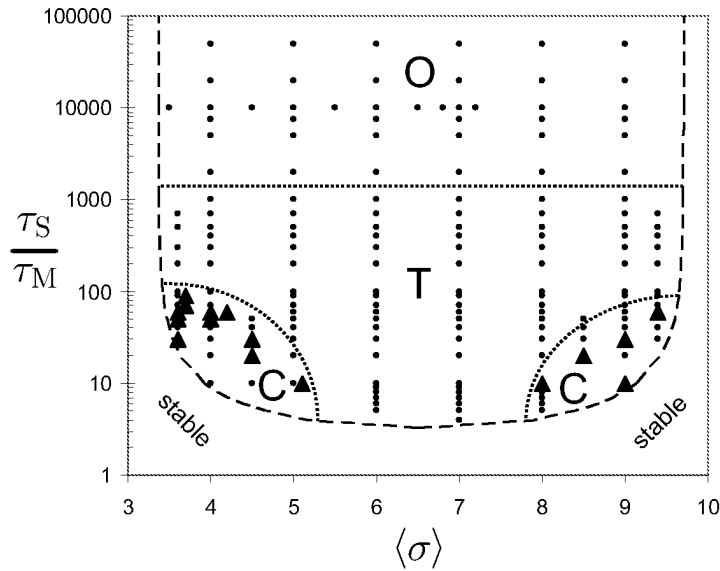}
\caption{\label{PhaseDiagram} Phase diagram of the model when $\taus$ and $\avstress$
are varied: ($\blacktriangle$)~chaotic point, ($\bullet$)~periodic
point. Three main regimes are observed: (O)~oscillating shear
bands, (T)~travelling shear bands, (C)~chaotic regions. The outer
dashed line is the linear stability limit,
$R'(\avstress)+1/\taus=0$ [\eq{InstabilityCondition}]. Numerical
parameters as in Fig.~\ref{FlowCurves}.}
\end{figure*}
To get a global picture of the model's behavior, we made a
systematic exploration of its spatio-temporal dynamics by varying
the two physically most important parameters: the ratio between
the structural timescale and the mechanical timescale
$\taus/\taum$, and the spatially averaged stress
\avstress (fixed by the mechanical torque on the Couette).
On varying $\taus/\taum$, we choose to keep $\taum$ fixed
($\taum=1/a=0.01$) and vary \taus only. All other parameters are
held at their values of Fig.~\ref{FlowCurves}, i.e., $b=20$,
$c=1.02$, $\lambda=40$, $\kappa=0.01$, $H=1$.

Despite the high-dimensional dynamical system under consideration
($2N=80$), the obtained `phase diagram' for the model (shown in
Fig.~\ref{PhaseDiagram}) displays a simple overall structure where
three main dynamical regimes emerge: periodic response with (more
or less complex) oscillating shear bands at extremely long
\taus; periodic response with travelling bands for long \taus; and
finally, chaotic response at shorter \taus and off-centered values
of \avstress. The dotted lines separating regions on the phase
diagram are guides to the eye, representing crossovers not sharp
transitions between different types of behavior. The `C-regions'
marked in the plot were defined so as to enclose all observed
chaotic points; these regions do, however, contain internal
structure with periodic and chaotic pockets, whose exact
boundaries can depend on initial conditions and other details.

We next discuss in some detail the three main regimes encountered
on the phase diagram, emphasising an intuitive understanding of
the physics involved. However, as shall be seen, there is
significant variety in behaviour even within the same regime.
\subsubsection{Periodically oscillating shear bands}
\begin{figure}
\includegraphics[width=.5\textwidth]{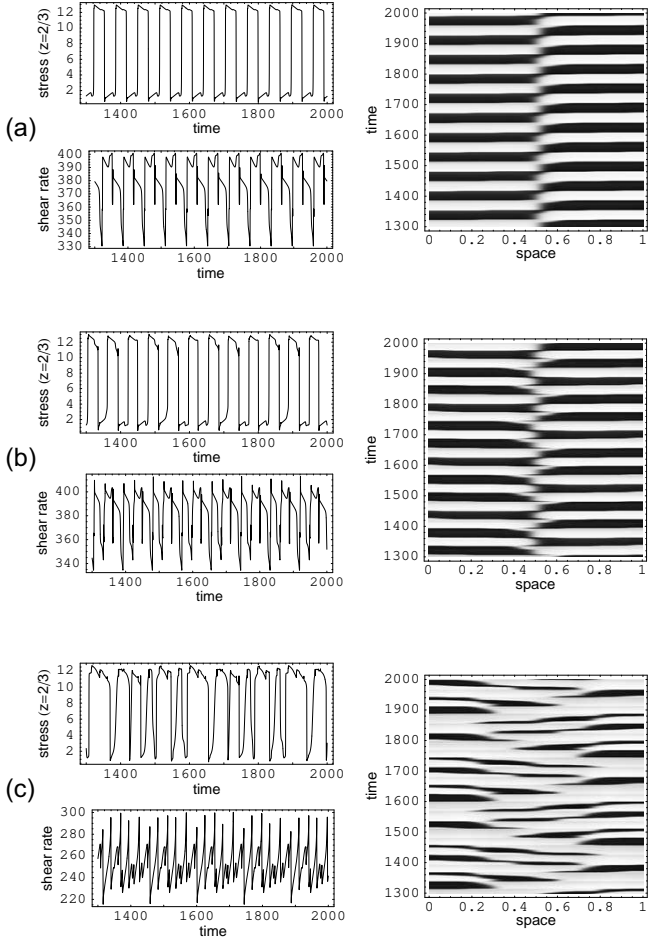}
\caption{\label{OscillatingBands}
Typical periodic responses in the oscillating shear band regime,
chosen along the line $\taus/\taum=10^4$ in the phase diagram.
(a)~Flip-flopping bands at $\avstress=7$; (b)~Zig-zagging
interface at $\avstress=7.1$; (c)~Complex periodic motion at
$\avstress=9$. Each group presents time series of the stress
$\sigma$ at $z=2/3$, the shear rate $\dot\gamma$, and a space-time
plot of $\sigma(z,t)$ with $t$ vertical, $z$ horizontal (clear
shades correspond to high stress, dark shades to low stress).
Parameters as in Fig.~\ref{FlowCurves}.}
\end{figure}

We first discuss the regime encountered when structural evolvution
is much slower than stress relaxation (typically $\taus/\taum
\geq 10^3$) marked `O' in Fig.~\ref{PhaseDiagram}. Here periodic
oscillations of the shear rate and stress are observed, while
spatially, the flow presents oscillating shear bands. Depending on
the imposed stress \avstress, the waveforms of these band
oscillations (which induce oscillations in both the local stress
$\sigma(z,t)$ and the mean strain rate $\dot\gamma(t)$) range from
simple to very complex. Moving along any horizontal line within
the region O (changing $\avstress$ at constant $\taus/\taum$), one
observes the same succession of behaviors; we present these for
one typical line, $\taus/\taum=10^4$.
\paragraph{`Flip-flopping' shear bands.}
\label{FlipFlopBands}
Near the middle of this line, e.g., $\avstress=7.0$, we find two
`flip-flopping' bands (see Fig.~\ref{OscillatingBands}-a): at any
time, the cell is equally divided between a high-stress and a
low-stress band, but the identity of the bands reverses
periodically. This results in the `checkerboard' spatiotemporal
pattern shown in Figure~\ref{OscillatingBands}-a. Accordingly, the
stress $\sigma(z,t)$, measured for a given height within the cell
(e.g., $z=2/3$), displays periodic oscillations with a waveform
close to a square wave (and abrupt changes between the low- and
high-stress states, as expected for `relaxation
oscillations'~\cite{Glendinning}). The flip-flop period $\tauf$ is
of order the structural time \taus. The shear rate $\dot\gamma(t)$
also presents a periodic evolution, albeit with a more complicated
waveform. (This extra complexity is generic in our shear rate time
series, but seems to be a model-dependent feature rather than
connected to any deep physics.)

The mechanism underlying the dynamics of flip-flopping bands is as
outlined in Sec.~\ref{OriginsInstability}: because of the
short-term instability present in the fluid
(Fig.~\ref{FlowCurves}), shear-bands form. Each of these bands is
locally submitted to a van-der-Pol-type instability which forces
them to oscillate between states of high and low stress. But these
local oscillations cannot occur independently: because the spatial
mean of the stress \avstress must be conserved, they have to be
synchronous---when one band goes up, the other must flip down.
Note that shear bands not only display different values of the
stress, but, in general also of the memory $m(z, t)$, albeit much
less markedly: in our model, this means that both the mechanical
and structural states differ between bands.

\begin{figure}
\includegraphics[width=.3\textwidth]{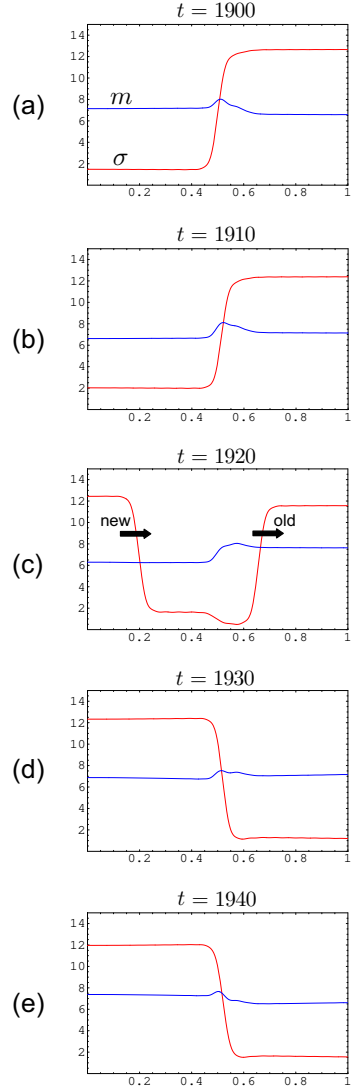}
\caption{\label{InterfaceReplacement}
Plots of $\sigma(z, t)$ and $m(z, t)$ vs. $z$ for successive
times, showing how the `old' interface is replaced with a `new'
one in the regime of flip-flopping bands. In (a) and (b) one sees
the latency period with stationary interface---note however that
$m$ slightly evolves from (a) to (b), thus preparing the
forthcoming flip; (c)~As the flip occurs, travelling wave
replacing the previous interface by a new one; (d) and (e)~new
latency period---note again the slow evolution in $m$.}
\end{figure}

Finally, a physically important question in this regime is: what
happens to the interface between bands during flip-flops? Looking
at Fig.~\ref{OscillatingBands}-a, we see that the interface
position seems stationary. A fixed position for the interface
would imply that, in the short interval when bands are flipping,
the interface profile has to reverse its slope. Interestingly,
closer scrutiny shows that this is \emph{not} what happens:
instead of reversing the slope of a static interface, the system
prefers to replace the `wrong' interface by a new one with the
correct slope. As seen in Fig.~\ref{InterfaceReplacement}, this
occurs by a rapid sweeping motion where the old interface quickly
travels to one end of the cell and disappears; simultaneously, the
new interface is generated at the other end and moves to the
middle of the cell, where it will rest until the next flip occurs.
Thus, long periods of immobility for the interface alternate with
rapid sweeping motions (the latter being too quick to be
observable on the scale of the plot in
Fig.~\ref{OscillatingBands}-a).

Qualitatively, everything happens as though when flips occurs in
the bands, they trigger a travelling wave removing the old
interface and bringing the new one in. Thus, this regime of
flip-flopping bands must be seen as one of intermittent travelling
waves, separated by long periods of latency (of order $\taus/2$).
In accord with this view, when the latency interval diminishes
sufficiently (i.e., \taus is decreased), the intermittent waves
should become continuous: this is indeed exactly what happens in
the `T'-region of the phase diagram (discussed in
Section~\ref{TravellingBands}).
\paragraph{Oscillating bands with zig-zagging interface.}
Pursuing our exploration of the model's behavior along the
horizontal line $\taus/\taum=10^4$, we now move slightly
off-center to the right or the left; for example, $\avstress=7.1$.
The interface between bands now adopts a periodic, zig-zagging
motion which superposes to the synchronous oscillations of the
shear bands (see space-time plot in
Fig.~\ref{OscillatingBands}-b). Accordingly, the time series of
the stress becomes more complex (square waves are distorted), and
the period becomes a multiple (here three) of $\tauf$. A similar
increase in complexity is seen in the shear rate.

The zigzag motion arises because the off-centered value of
\avstress now enforces unbalanced proportions of the
low- and high-shear bands in the Couette cell. Because this value
for \avstress is fixed, these proportions are also fixed: thus
each time shear bands oscillate and reverse identities, the
interface between them must move to and fro to maintain the
required proportions.
\paragraph{\label{ComplexOscillations}Complex periodic oscillations.}
If we now move further towards the wings of the phase diagram, the
coupling between the global constraint on \avstress (unequal
bands) and the intrinsic flipping dynamics of the bands generates
an extremely complex behavior in the fluid, which, remarkably,
manages nonetheless to remain periodic.
Figure~\ref{OscillatingBands}-c shows the response obtained at
$\avstress=9.0$: the time series for the stress $\sigma(2/3,t)$,
though still related to the simple oscillations seen at
$\avstress=7.0$, has become very ragged, and the period is now a
large multiple of $\tauf$ (about six times).

We conclude this presentation of regime O by noting that, in
experiments, oscillating shear bands similar to our theoretical
results have indeed been observed in shear-thickening
fluids~\cite{Wheeler,Fischer1,Fischer2,Fischer3}.
\subsubsection{Travelling shear bands\label{TravellingBands}}
\begin{figure}
\includegraphics[width=.5\textwidth]{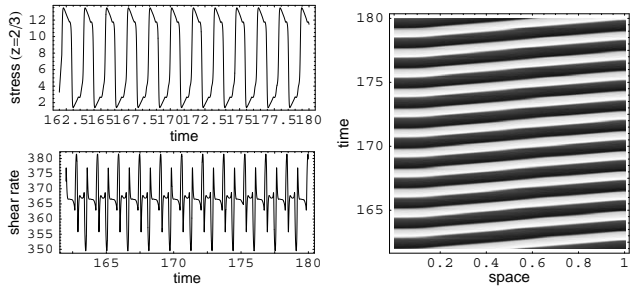}
\caption{\label{TravellingFig}
Periodic time series and space-time plot of the stress
$\sigma(z,t)$ in the travelling band regime. Parameters:
$\taus/\taum=90$, $\avstress=7$, others as in
Fig.~\ref{FlowCurves}.}
\end{figure}

We now describe the second regime in the phase diagram of our
model, encountered when \taus is long as compared to \taum , but
not exceedingly so ($10\leq\taus/\taum\leq10^3$; region~`T' in
Fig.~\ref{PhaseDiagram}). This regime is characterised by a
periodic nucleation of shear bands which subsequently cross the
system at roughly constant velocity. The typical situation is
shown in Fig.~\ref{TravellingFig}: nucleation occurs at a
boundary, and bands travel across the entire system, one at a
time. As before, the period of the various time series is
comparable to the structural timescale \taus.
Similar band motion was reported in Ref.~\onlinecite{Fielding},
but there the bands `ricochet' off the walls of the container.

\begin{figure}
\includegraphics[width=.4\textwidth]{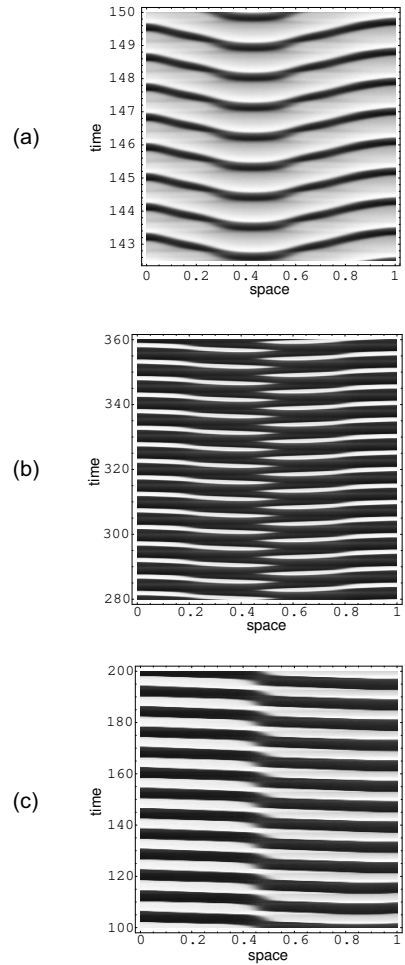}
\caption{\label{TravellingExamples}
Space-time plots of the stress $\sigma(z,t)$ showing other
examples of travelling bands. (a)~`Target pattern' analogue seen
at $\taus/\taum=30$ and $\avstress=8.5$; (b)~Alternating bands
seen at $\taus/\taum=400$ and $\avstress=5$; (c)~Transition
between oscillating band and travelling band regime seen at
$\taus/\taum=1000$ and $\avstress=7$. Other parameters as
previously.}
\end{figure}

Figure~\ref{TravellingExamples} shows examples of other travelling
behaviors in this regime: The first space-time plot~(a) presents a
case where nucleation occur at a non-boundary point, yielding two
outgoing waves in a one-dimensional analogue of the classical
`target patterns' seen in chemical
oscillators~\cite{ChemicalOscillators}. Space-time plot~(b)
presents another case of interior nucleation, but where bands
alternatively travel to one edge or the other. (Details here
strongly depend on the initial conditions.) Finally, space-time
plot~(c) illustrates how the model behavior crosses over smoothly
from oscillating bands in regime O to travelling bands in regime
T: the behavior is intermediate between the situations depicted in
Fig.~\ref{OscillatingBands}-a and Fig.~\ref{TravellingFig}, with
bands somehow oscillating as they travel.

Note that the waves observed in regime T are kinematic waves,
arising from a staggered phase distribution in the local
van-der-Pol band oscillations. They are not associated with
transport of material or stress; hence the waves can travel
through the system boundaries, despite the imposed boundary
conditions ($\nabla\sigma=0$). Accordingly the wave velocity is
independent of the stress diffusion constant $\kappa$ (data not
shown).

\begin{figure}
\includegraphics[width=.4\textwidth]{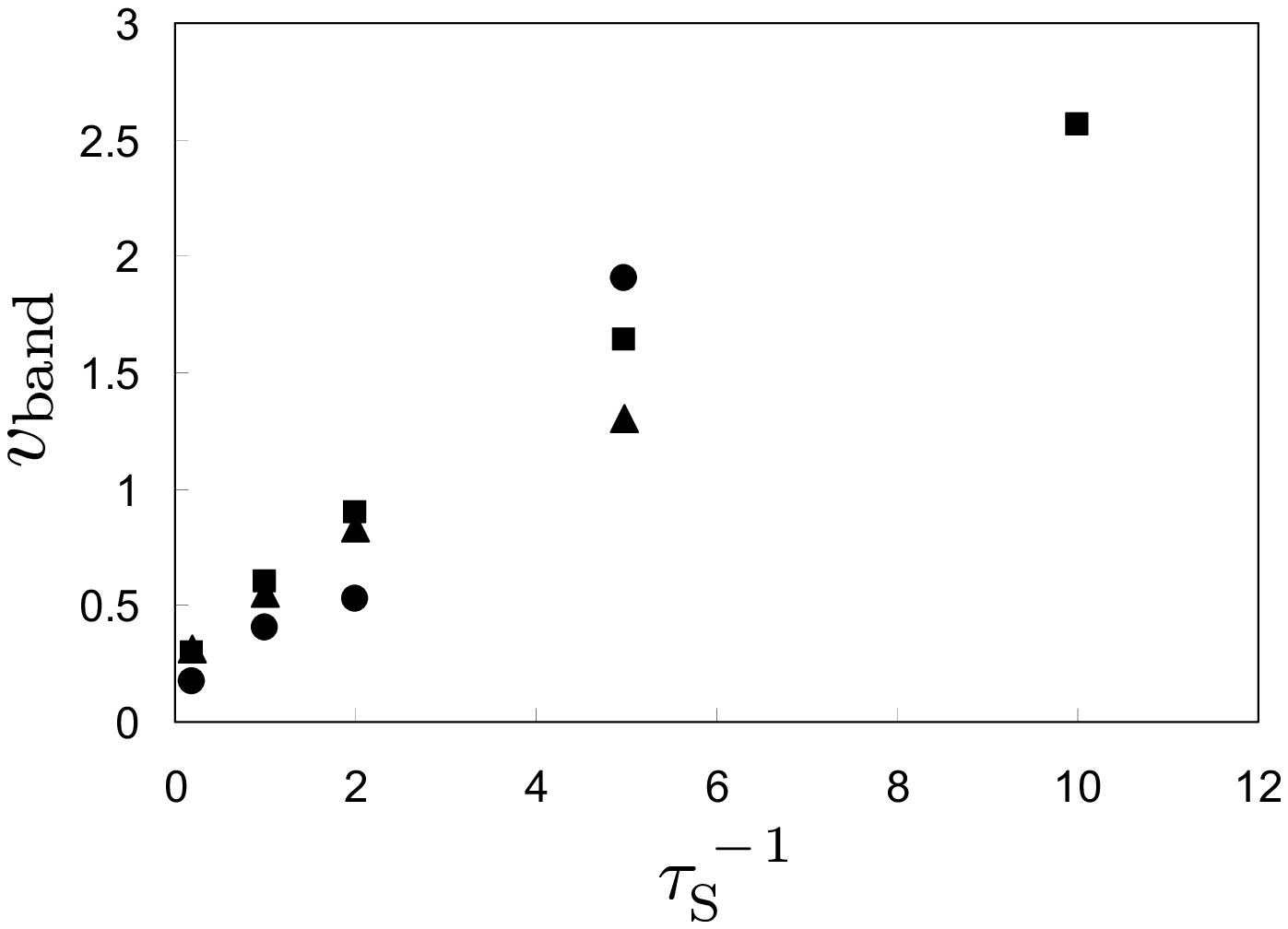}
\caption{\label{BandVelocity}
Velocity of travelling bands vs.\ $\taus^{\;-1}$, measured in
situations where a single band is present in the system:
$(\displaystyle\bullet)~\avstress=8.0$;
$({\scriptstyle\blacksquare})~\avstress=7.0$;
$(\blacktriangle)~\avstress=6.0$. The plot shows that
$v_\text{band}$ is roughly proportional to $\taus^{-1}$. Error
bars are $\pm 0.15$. Parameters as in previous figures.}
\end{figure}

Finally, we studied how the bands velocity varied as \taus was
changed. At least for the situation in Fig.~\ref{TravellingFig},
where there is only one band at a time in the cell, a naive
argument for the velocity is as follows. Since travelling bands
correspond to (staggered) local oscillations, two successive
passages of bands at a given point of the system correspond to the
completion of a local oscillation cycle for this point. Noting
that the condition of a constant \avstress imposes that bands
disappearing at one end must reappear immediately at the other, we
conclude that each band must cross the system in one local
oscillation period. Since this is of order \taus, and recalling
that $H=1$, we deduce $v_\text{band} \sim 1/\taus$ for the band
velocity. Our data show that $v_\text{band}$ indeed decreases with
\taus, and confirm a roughly linear trend with $\taus^{-1}$ (see
Fig.~\ref{BandVelocity}).
\subsubsection{Spatiotemporal rheochaos\label{Rheochaos}}
\begin{figure*}
\includegraphics[width=\textwidth]{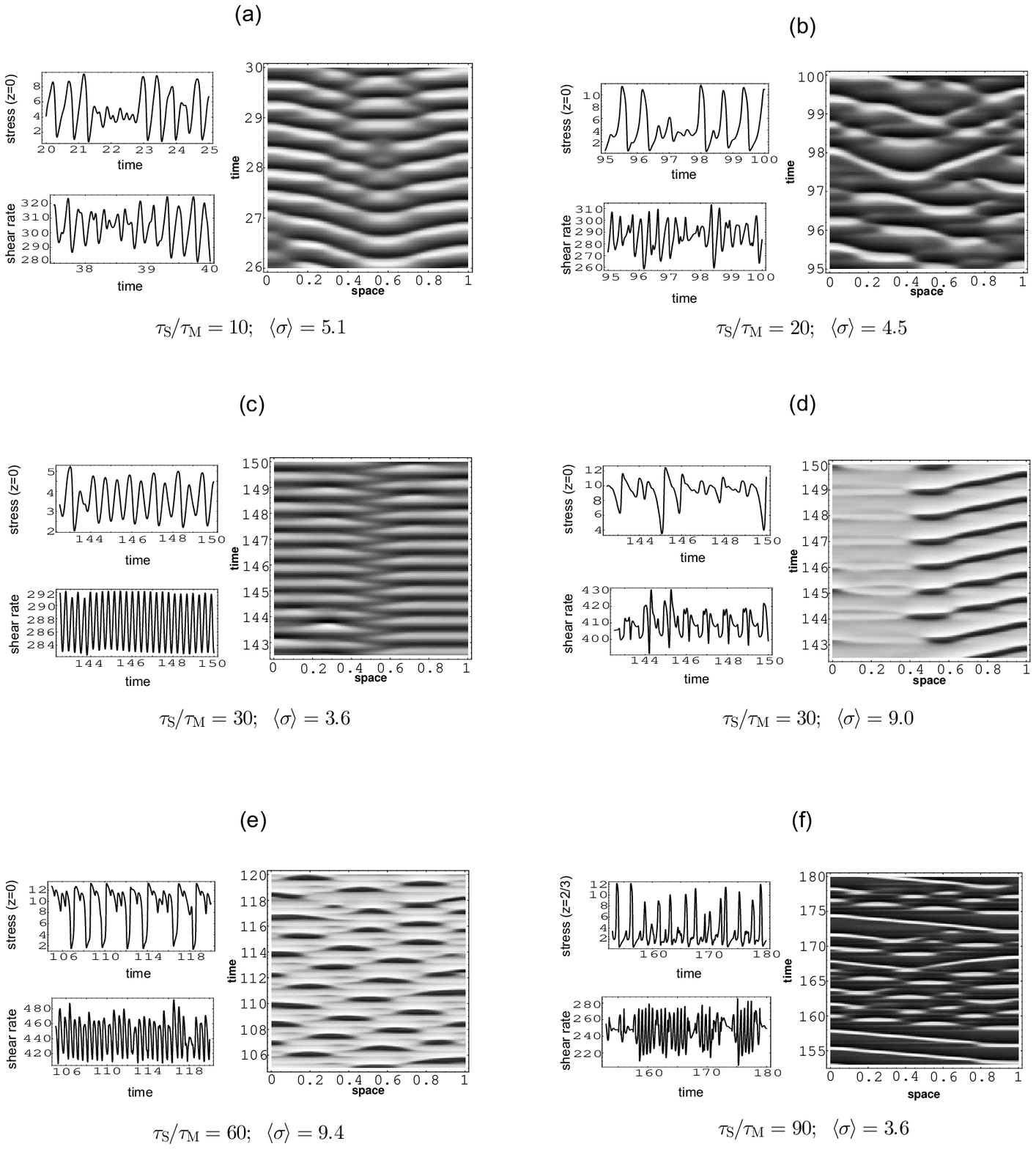}
\caption{\label{RheochaosExamples}
Various types of spatiotemporal rheochaos observed in the model
(high-order truncation). Numerical parameters as in
Fig.~\ref{FlowCurves}.}
\end{figure*}
The third main regime encountered in the phase diagram arises in
two disconnected pockets (regions `C' in Fig.~\ref{PhaseDiagram})
at moderately long values of \taus relative to \taum and for
strongly off-centered values of the imposed stress \avstress. An
increased complexity of oscillations within regime O on
approaching the wings of the phase diagram was already noted
previously; in regions C, it finally leads to chaotic behavior.

The spatiotemporal patterns produced in this regime are extremely
varied and often appear like complex versions of periodic patterns
that arise in other parts of the phase diagram. Examples of
rheochaos are given in Figure~\ref{RheochaosExamples}: in (a), we
have a regime of chaotic bands with `random defects' between bands
(this is similar to rheochaotic patterns observed in
shear-thinning micelles in Ref.~\onlinecite{Fielding}); in (b), we
see a situation which resembles the travelling bands of regime T,
but here disordered and following a somewhat wiggling motion; plot
(c) presents the case of flip-flopping bands where irregularities
appear within the bands themselves; plot (d) shows travelling
bands whose nucleation point changes `at random'; in (e) and (f),
we observe a `bubbly' phase of localized, short-lived shear bands
that appear and disappear erratically in the cell (although
occasionally, a band survives and shoots across the system).

\begin{figure}
\includegraphics[width=.4\textwidth]{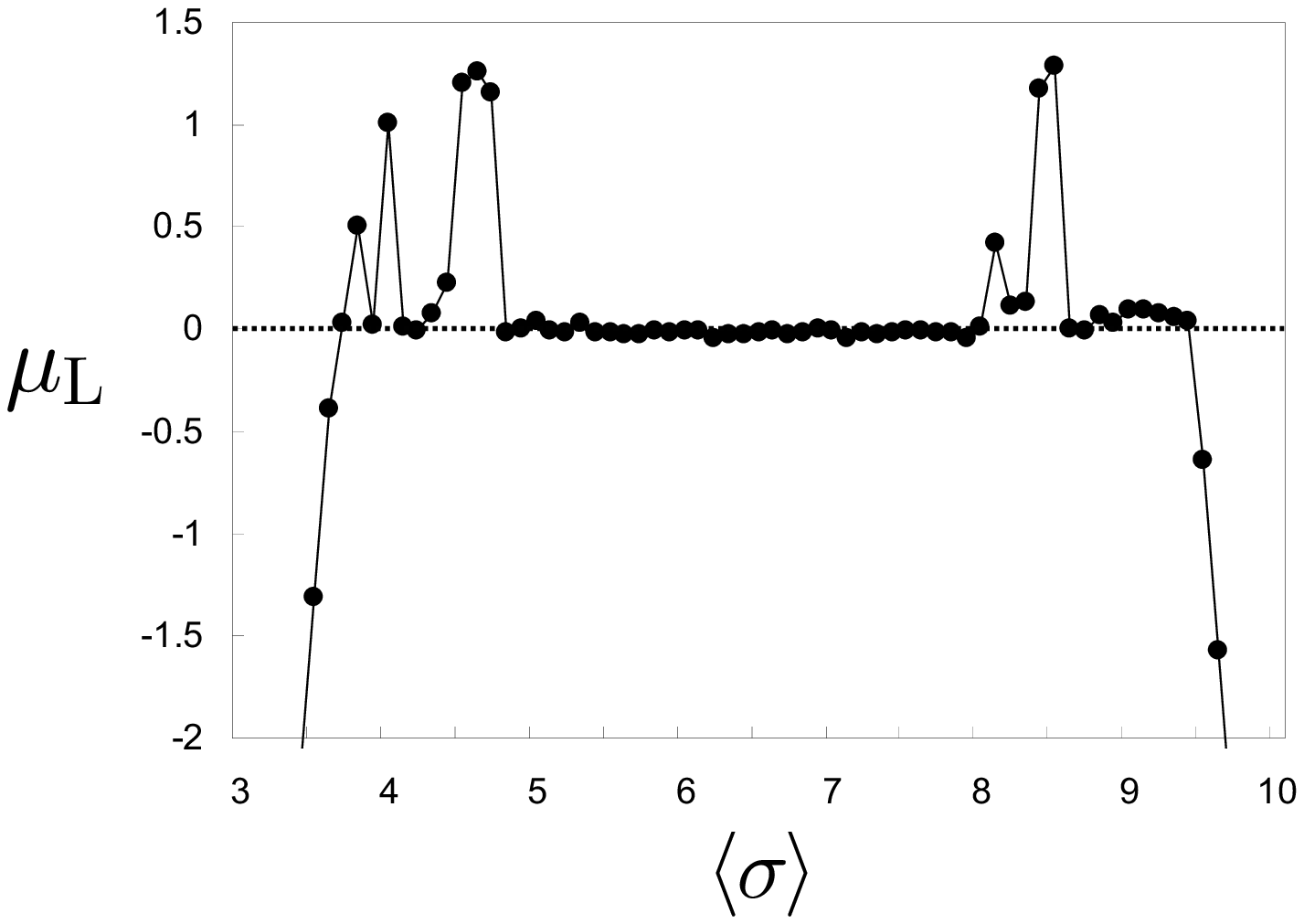}
\caption{\label{Lyapunov}
Plot of the largest Lyapunov exponent $\mu_\text{L}$ computed as a
function of the imposed stress
\avstress on the line $\taus/\taum=20$ (high-order truncation). Chaotic behavior (positive
values) appears only on the wings of the diagram. Imprecision on
the exponent values is typically $\pm0.1$. Numerical parameters as
in Fig.~\ref{FlowCurves}.}
\end{figure}

Chaos, for each point so marked in Fig.~\ref{PhaseDiagram}, was
confirmed by computing the largest Lyapunov exponent,
$\mu_\text{L}$. In Fig.~\ref{Lyapunov}, we give a plot of
$\mu_\text{L}$ when \avstress is varied on the horizontal line
$\taus/\taum=20$ of the phase diagram.

Returning to the data of Fig.~\ref{RheochaosExamples}, these
patterns vary in how highly developed is the chaos: for instance,
pattern~(b) seems much more erratic than pattern~(c). This is
reflected on the Lyapunov exponents: we find $\mu_\text{L}\simeq
1.2$ and $\mu_\text{L}\simeq 0.2$ respectively. However, we were
unable to demonstrate true low-dimensional chaos within our
high-order truncation results. Nor could we numerically explore
any route to chaos, as the transition window proved unattainably
small in the phase diagram. We do however discuss below the route
to chaos within the low-order truncation of the model.

A last comment regards the role played by the value of $\kappa$,
the stress diffusivity, in chaotic regimes. Numerical solutions in
this article were computed with a default value of
$\kappa=10^{-2}$, but data obtained with different values suggest
that smaller values of the stress diffusivity favor chaos. This
makes sense, as $\kappa$ is a damping term for higher Fourier
modes [see \eq{ShortModeStressEq}]: when $\kappa$ is smaller, more
modes are involved in the dynamics, and thus the system is
effectively higher-dimensional. Realistic values for $\kappa$
might be much smaller than our choice, which was dictated by the
need to find $N$-independent behavior at relatively modest $N \ge
25$. Accordingly a more realistic phase diagram at lower $\kappa$
might display much larger chaotic pockets than
Fig.~\ref{PhaseDiagram}.
\section{Low-order results\label{lowtruncation}}
The results presented so far all used a high-order truncation,
with a view to obtaining a numerically reliable and precise
representation of the continuum model of
Eqs.~(\ref{RheologicalEquation})--(\ref{StructuralEquation}). In
this section, we instead study the behavior in the simplest
nontrivial mode decomposition of
Eqs.~(\ref{RheologicalEquation})--(\ref{StructuralEquation}), with
$N=3$. The corresponding low-dimensional results provide only a
caricature of the full dynamics; but they bring interesting
insights. This approach is similar in spirit to the classical
simplification of the Rayleigh-B{\'e}nard equations (describing
thermal convection) into the Lorentz
equations~\cite{MannevilleLivre1}.
\begin{figure*}
\includegraphics[width=.65\textwidth]{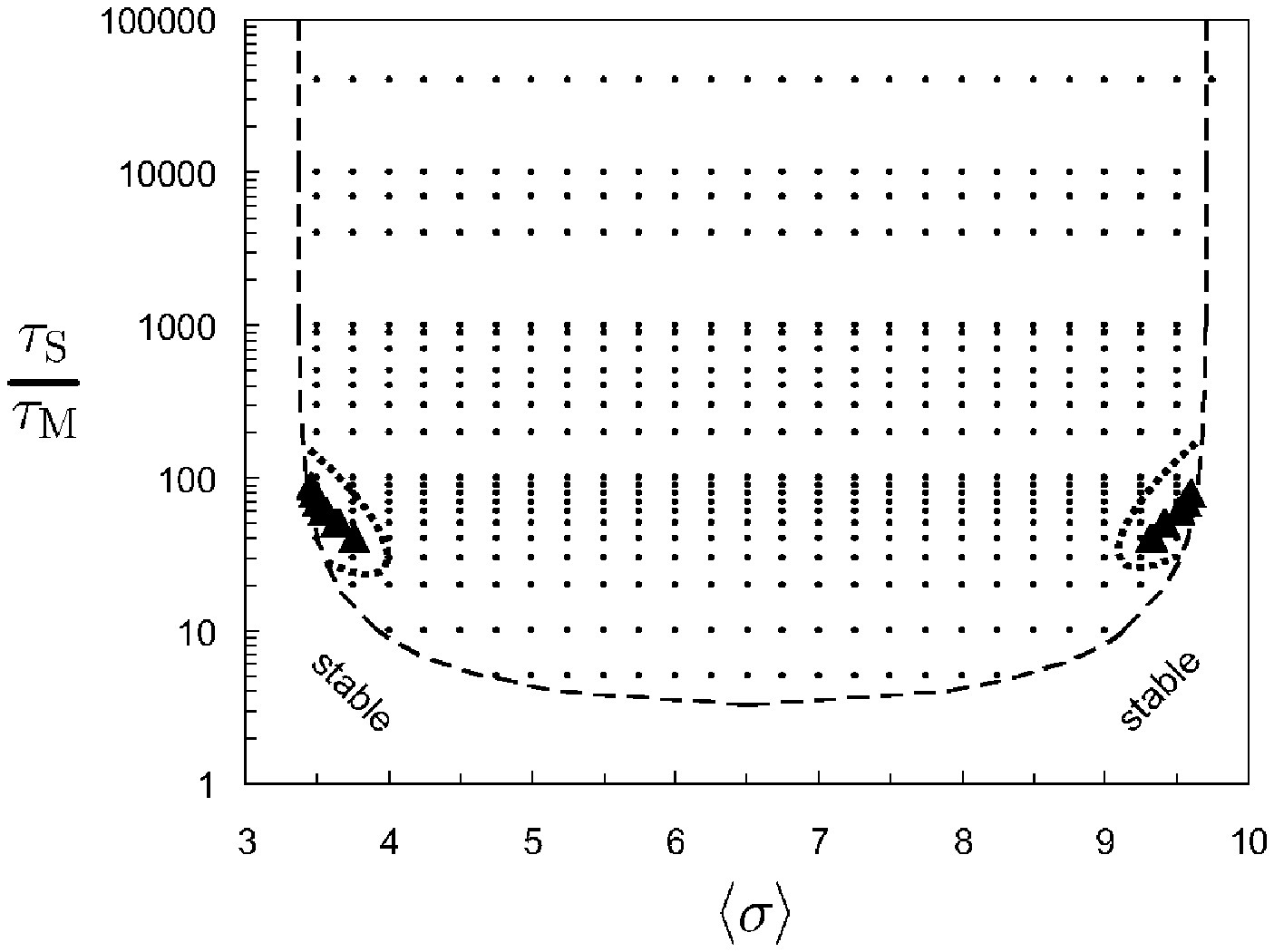}
\caption{\label{PhaseDiagramLow}
Phase diagram in the low-dimensional truncation of the model:
($\bullet$)~Periodic points; ($\blacktriangle$)~Chaotic points.
Note that, for pictural clarity, a horizontal spacing between
points $\Delta\avstress=0.25$ was chosen for the plot, but our
actual numerical exploration of the phase diagram was performed on
a much finer grid with $\Delta\avstress=10^{-2}$. The outer dashed
line delineates the limit of linear stability
[\eq{InstabilityCondition}]. Dotted lines around chaotic pockets
are merely guides to the eye. Numerical parameters as in
Fig.~\ref{FlowCurves}.}
\end{figure*}

By restricting the number of dynamical variables, it permits  an
analytical or semi-analytical study of some of the model's
features. It also clarifies which features of the model survive in
low dimension and thus are in some sense robust; we shall see that
rheochaos survives the truncation, and is not therefore solely the
result of high-order couplings in the full dynamics.

Finally, there are interesting physical consequences to
demonstrating the presence of chaos in a low-dimensional Fourier
decomposition. The results we have presented so far, along with
other works~\cite{CatesHead,Fielding,Ramaswamy}, might be taken to
suggest a generic interpretation of rheochaos in complex fluids as
resulting from the erratic motion of discrete interfaces between
shear bands. This idea is appealing because, in turn, it suggests
that rheochaos may be more efficiently modelled by focussing on
interfacial dynamics and writing an equation of motion directly
for the interface~\cite{CatesHead,AjdariInterface} (rather than
for the whole fluid as we are doing here). The presence of
rheochaos in our low-order truncation (which allows only smooth
spatial variations of $\sigma$ and $m$) shows that sharply defined
bands are not an automatic prerequisite for rheochaos.

As already explained in Sec.~\ref{numerics}, the lowest possible
mode decomposition allowing chaos is obtained for $N=3$. Since the
uniform Fourier modes, $\sigma_0=\avstress$ and $m_0$, are
dynamically inactive, this reduced representation of the model is
effectively four-dimensional, with $\sigma_1(t)$, $\sigma_2(t)$,
$m_1(t)$ and $m_2(t)$ as the degrees of freedom. Drawing from the
general analytical expression given in
Appendix~\ref{FullAnalytical}, the dynamical equations for the
four retained modes have the relatively simple form (recall
\avstress is a constant):
\begin{eqnarray*}
\dot\sigma_1&=&  \Bigl[- a +
2 b \avstress - 3 c \avstress^2 -
\kappa\Bigl(\frac{\pi}{H}\Bigr)^2\,\Bigr] \sigma_1 - \frac{3}{4} c\, \sigma_1^{\,3}\\ &&
+ \bigl[b-3c\avstress\bigr] \sigma_1  \sigma_2 - \frac{3}{2} c\,
\sigma_1 \sigma_2^{\,2} -\lambda m_1\\
\dot\sigma_2&=& \Bigl[ - a  + 2 b \avstress - 3 c \avstress^2
-\kappa\Bigl(\frac{2\pi}{H}\Bigr)^2 \Bigr]\sigma_2 - \frac{3}{4} c
\,\sigma_2^{\,3} \\&&+
\frac{1}{2}\bigl[b- 3c \avstress\bigr]\sigma_1^{\,2} -
\frac{3}{2} c\,\sigma_1^{\,2} \sigma_2  -\lambda m_2 \\
\dot m_1&=&(\sigma_1-m_1)/\taus\\
\dot m_2&=&(\sigma_2-m_2)/\taus~.
\end{eqnarray*}

As with the high-order truncation, we have explored numerically
the phase diagram for this minimal model. We indeed find chaos,
confirming that rheochaos is a robust feature of the underlying
physics. As Figure~\ref{PhaseDiagramLow} shows, the global
structure of the phase diagram is also robust and presents the
same features as in the high truncation: chaotic behavior appears
on both wings of the diagram (off-centered values of
\avstress) and for $\taus/\taum$ ratios not too large (structural
relaxation not too slow), while periodic states are observed
elsewhere. One observation is that chaotic pockets are now much
smaller than in the high-order truncation phase diagram: this is
natural as chaos is usually facilitated in higher dimensions.

\begin{figure}
\includegraphics[width=.5\textwidth]{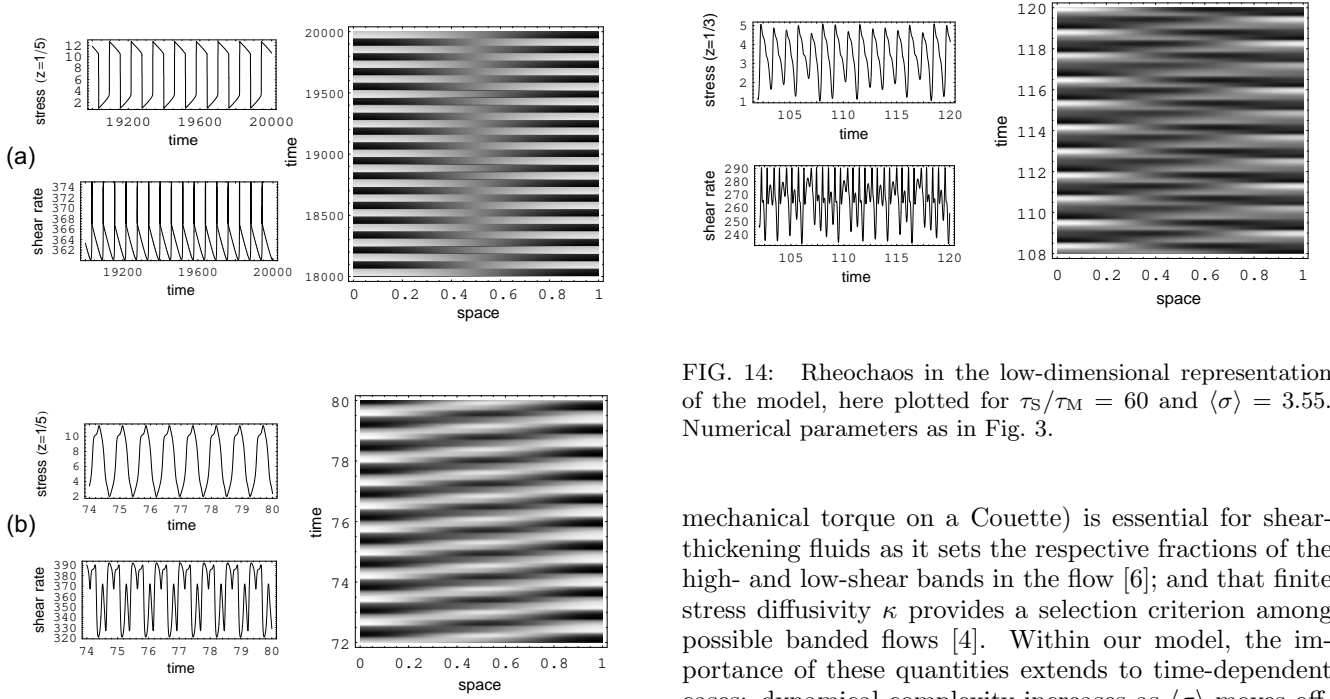}
\caption{\label{LowDimAnalogues}
Low-dimensional analogues to the oscillating band regime~(a) and
to the travelling band regime~(b). Note the ill-defined
`interface' between `bands'. Parameters: (a)~$\taus/\taum=10^4$,
$\avstress=7$; (b)~$\taus/\taum=40$, $\avstress=7$; other
parameters as in Fig.~\ref{FlowCurves}.}
\end{figure}

We do not classify periodic points in this low-order model as
`oscillating bands' or `travelling bands': the stress variations
are anyway too smooth to allow the definition of proper shear
bands. However, as illustrated in Fig.~\ref{LowDimAnalogues},
there remain low-dimensional analogues of the oscillating and
travelling bands, in roughly the same locations as in the
high-order truncation.
\begin{figure}
\includegraphics[width=.5\textwidth]{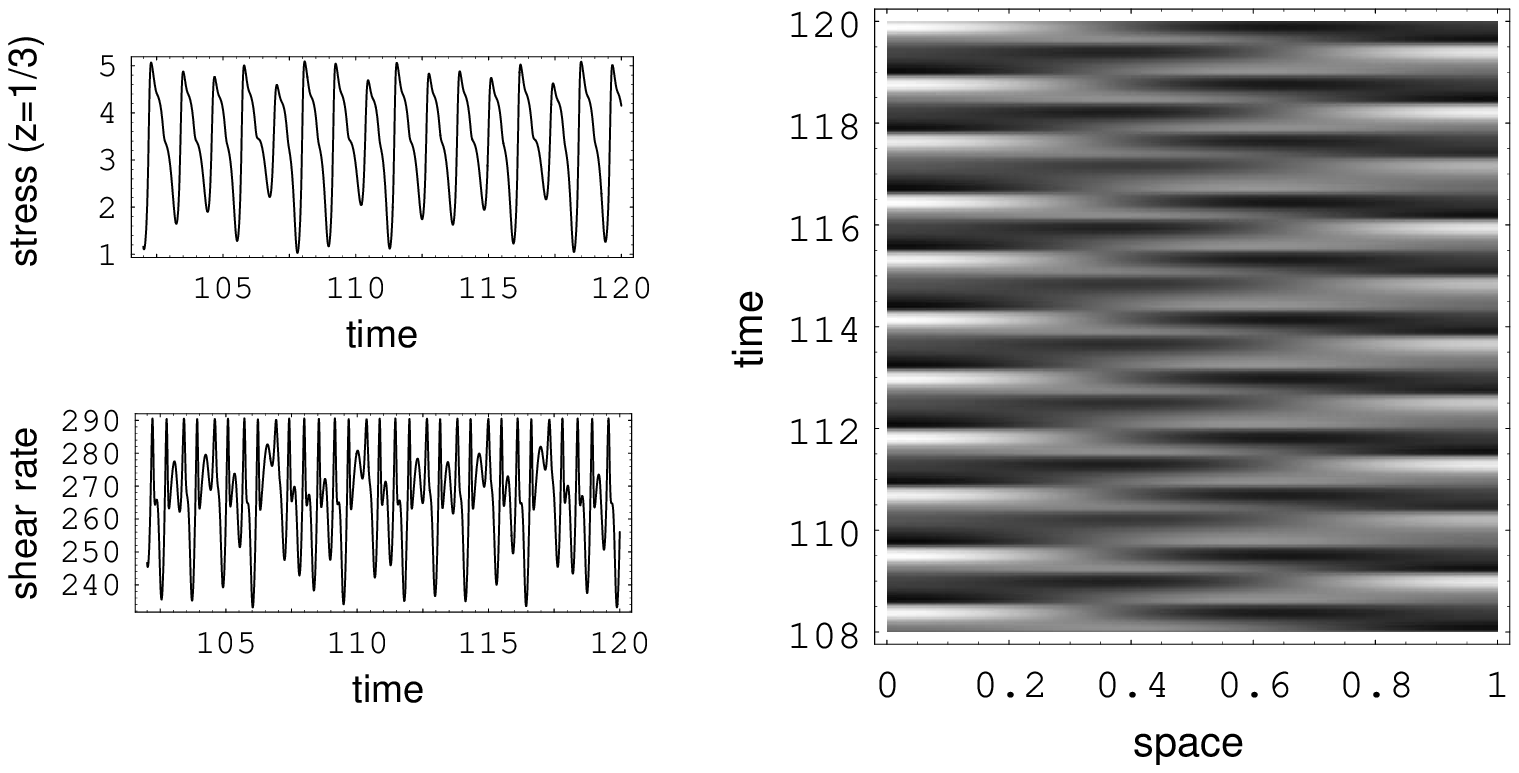}
\caption{\label{LowDimChaos}
Rheochaos in the low-dimensional representation of the model, here
plotted for $\taus/\taum=60$ and $\avstress=3.55$. Numerical
parameters as in Fig.~\ref{FlowCurves}.}
\end{figure}
In Figure~\ref{LowDimChaos}, we show a typical trajectory in the
chaotic regime of the low-order model: this appears less erratic
than in the high-order counterpart (Fig.~\ref{RheochaosExamples}),
essentially amounting to a slightly irregular oscillation of two
`bands'. The largest Lyapunov exponent is $\mu_\text{L}\simeq
0.4$. Finally, we studied the route to chaos in the low-order
truncation, and found a classical period-doubling scenario
(Fig.~\ref{RouteChaos}). This result does not, however, imply
anything about the route to chaos in the high-dimensional version
of the model.
\begin{figure}
\includegraphics[width=.45\textwidth]{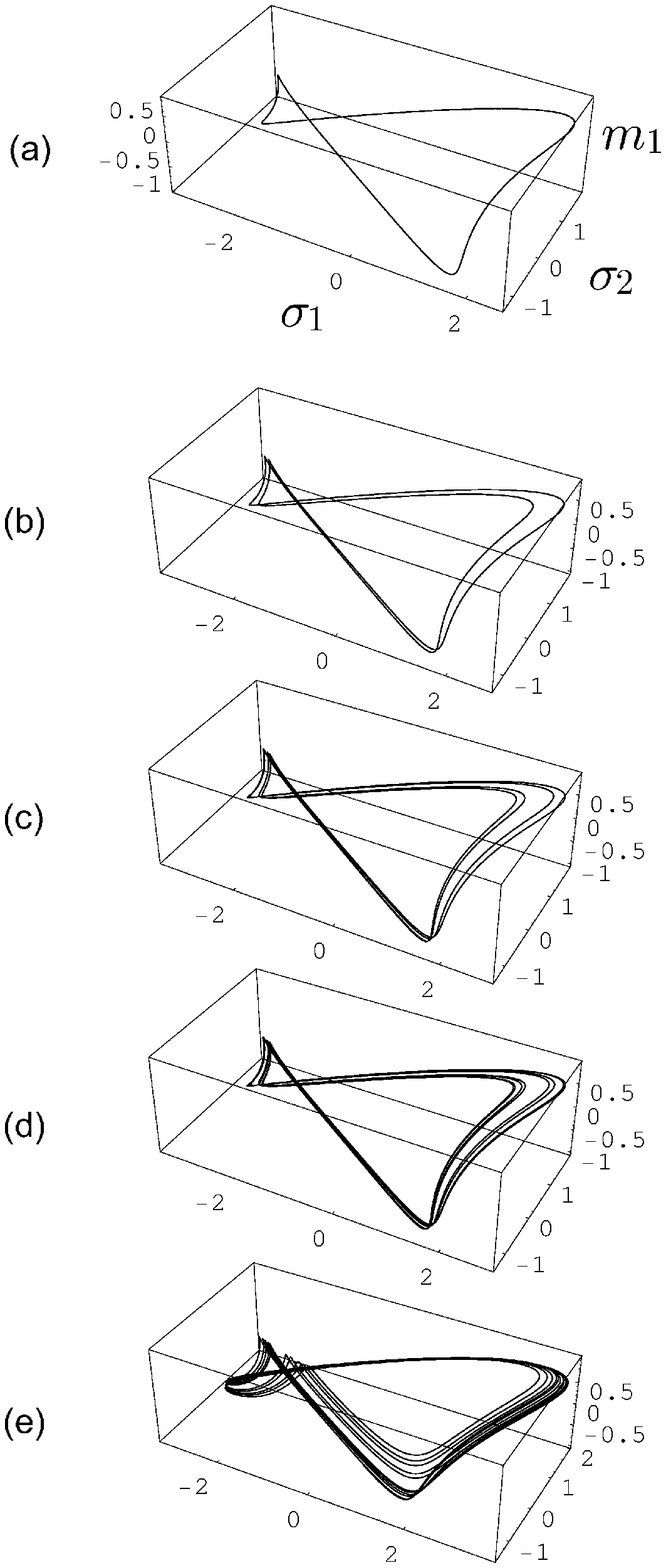}
\caption{\label{RouteChaos}
Route to chaos in the low-dimensional truncation, shown for
$\taus/\taum=60$. Plotted are three-dimensional projection on the
$(\sigma_1,\sigma_2,m_1)$-space of the four-dimensional
attractors. From (a) to (d), first steps of the period-doubling
cascade: (a)~$\avstress=3.53$, (b)~$\avstress=3.535$,
(c)~$\avstress=3.5375$, (a)~$\avstress=3.5379$; (e) Chaotic
attractor obtained for $\avstress=3.55$, corresponding to the
situation shown in Fig.~\ref{LowDimChaos}. Numerical parameters as
in Fig.~\ref{FlowCurves}.}
\end{figure}

The results of this section show, not only that sharp interfaces
are not necessary for rheochaos, but also that the low-dimensional
representation of our shear-thickening model can reproduce most of
the important physical features observed in higher dimensions.
This low-order truncation could also enable a deeper understanding
of the oscillating and travelling band regimes: we have not
explored in detail the relation between \tauf,
\taus, and the band velocity. It might also permit a deeper understanding
of the effect of the global stress constraint, particularly the
mechanism whereby this affect the dynamics and makes it more or
less complex.
\section{Conclusions}
\label{concsec}
%
%
%
%
We have introduced a shear-thickening fluid model, allowing for
spatial inhomogeneity, in which the relaxation of stress couples
to a  memory term linked to slow structural changes in the fluid.
The interplay between the rheological and the structural dynamics
leads to complicated spatio-temporal dynamics. The model maps onto
the FitzHugh-Nagumo model for neural networks, but with a
different type of global constraint (imposition of the average
stress \avstress), specific to complex fluids.

In steady state for both the structure and stress, the flow curve
is monotonically increasing. However, short-term flow curves,
valid on timescales too short for the structure to relax, are
non-monotonic and unstable. This obliges the appearance of
unsteady shear bands.

The phase diagram of the model on varying \avstress and the ratio
of structural and stress relaxation times, $\taus/\taum (\ge 4)$
has a simple overall structure. Three main regimes were found. For
very large $\taus/\taum$, we found oscillating shear bands;
elsewhere we found periodic travelling shear bands at mid-range
values of \avstress, with regions of spatio-temporal rheochaos for
off-centre values.

In the oscillating band regime, shear bands must oscillate
synchronously while respecting the imposed value for the mean
stress \avstress. When the volume fractions of the two bands
within the cell are unequal (off-centre \avstress) oscillations
become increasingly complex but remain periodic. In the other
periodic regime seen in the model, bands nucleate at a border
point of the container, or more rarely at an interior point, and
travel across it at a near-constant velocity which scales roughly
as $\taus^{-1}$.

Our model exhibits spatio-temporal chaos. The corresponding
space-time patterns are varied, but fall into a common picture
(coherent with that of other groups~\cite{Fielding,Ramaswamy}):
rheochaos manifests as a flow that restlessly attempts to form
steady shear bands, but fails due to internal structural
constraints.

It is known that the constraint on \avstress (set by the
mechanical torque on a Couette) is essential for shear-thickening
fluids as it sets the respective fractions of the high- and
low-shear bands in the flow~\cite{OlmstedGoveas}; and that finite
stress diffusivity $\kappa$ provides a selection criterion among
possible banded flows~\cite{LuOlmstedBall}. Within our model, the
importance of these quantities extends to time-dependent cases:
dynamical complexity increases as \avstress moves off-centre,
whereas a smaller diffusivity $\kappa$ promotes chaos.

Taking advantage of our Fourier-space representation, we also
studied a four-dimensional truncation for which we showed that the
important physical features found in high dimension persist,
including the general structure of the phase diagram.

The presence of chaos in the low-order model shows that sharp
interfaces between bands are not necessary for rheochaos. This
does not mean that such interfaces have no effect, only that the
chaos would persist without them.

In conclusion then, our work, along with
others'~\cite{Fielding,Ramaswamy}, supports a `frustrated
shear-banding' picture of rheochaos, in which flow inhomogeneity
(whether sharp interfaces or smooth variations) plays a crucial
role. A gap between theory and experiment still remains: our model
makes no direct connection with the microscopics of the considered
fluids. Improved experimental information about the dimensionality
of the observed chaos, or the route into, for shear-thickening
systems would be most helpful. One general prediction of our model
that may be easily tested experimentally is that the dynamics
should get more complex as the imposed stress $\avstress$ is
chosen closer to the edges of the fluid's unstable window.

One potentially important factor that has been neglected in the
present model is the role played by normal stresses in the fluid.
In many of the experimental fluids we are concerned with, their
magnitude is significant in high-shear regimes. One could also
consider alternative scenarios for rheochaos, where stress
variations do not couple to nonconserved structural changes as
proposed here (these include modulation of concentration at finite
wavevector, i.e. microphase separation) but instead to conserved
modes such as global concentration fields.
\begin{acknowledgments}
The authors thank L.~B{\'e}cu, A.~Colin, S.~Fielding,
S.~Manneville, P.~Olmsted, D.~Roux and J.-B.~Salmon for
discussions. AA funded by EPSRC grant GR/R95098.
\end{acknowledgments}
\appendix
\section{Non-dimensionalisation}
\label{NonDimensional}
We here describe how the non-dimensional
equations~(\ref{spaceCHA})--(\ref{defRandM}) for the model are
obtained from the original physical equations. This is especially
useful if one wants to deduce the numerical value of a
non-dimensional parameter used in the model from an experimental
value. Throughout this Appendix, we use the convention that normal
letters are for dimensional, physical quantities, and starred
letters for non-dimensional ones. In the rest of the article, all
quantities are non-dimensional by default, so that stars have
systematically been omitted.

The dimensional version of \eq{spaceCHA} writes
\begin{eqnarray}
\label{DimensionalSpaceCHA}
\dot\sigma&=&G\dot\gamma-\taum^{-1}\sigma+b\sigma^2-c\sigma^3\\
\nonumber & &-\lambda\int\taus^{-1}e^{(t-t')/\taus}\,\sigma(t')\,\dd t' +
\kappa\nabla^2_{\! z}\sigma
\end{eqnarray}
where G is the transient elastic modulus, and other quantities
have been defined in the main text.

We use $G$, $\tau_0$ (defined below) and $H$ (the Couette axial
extent) as, respectively, the units for stress, time and space,
and define the corresponding reduced quantities:
$\sigma^\ast=\sigma/G$, $t^\ast=t/\tau_0$ and $z^\ast=z/H$.

Substituting these quantities in \eq{DimensionalSpaceCHA}, we
obtain the non-dimensional expression
\begin{eqnarray}
\label{NoDimensionalEq}
\dot\sigma^\ast&=&\dot\gamma-\frac{\tau_0}{\taum}\sigma^\ast
+bG\tau_0\sigma^{\ast 2}-cG^2\tau_0\sigma^{\ast 3}\\
\nonumber&
&-\lambda\tau_0^2\!\int\taus^{-1}e^{\tau_0(t^\ast-t'^\ast)/\taus}
\,\sigma^\ast(t'^\ast)\,\dd t'^\ast+
\kappa\frac{\tau_0}{H^2}\nabla^2_{\!z^\ast}\sigma^\ast
\end{eqnarray}

We now have to choose a value for the unit time $\tau_0$: for
practicality, because we always work on timescales longer than
$\taum$, we choose $\tau_0=100\taum$. We next define the reduced
Maxwell time $\taum^\ast=\taum/\tau_0$ and the constant $a^\ast$,
its inverse: we have $\taum^\ast=0.01$ and $a^\ast=\taum^{\ast
\;-1}=100$.

Finally, we transform the remaining physical parameters of the
model into reduced quantities: $b^\ast=bG\tau_0$,
$c^\ast=cG^2\tau_0$, $\lambda^\ast=\lambda\tau_0$,
$\taus^\ast=\taus/\tau_0$, $\kappa^\ast=\kappa\,\tau_0/H^2$. In
terms of these reduced quantities, Equation~\ref{NoDimensionalEq}
rewrites
\begin{eqnarray*}
\dot\sigma^\ast&=&\dot\gamma-a^\ast\sigma^\ast
+b^\ast\sigma^{\ast 2}-c^\ast\sigma^{\ast 3}\\
& &-\lambda^\ast\!\int\taus^{\ast
-1}e^{(t^\ast-t'^\ast)/\taus^\ast}
\,\sigma^\ast(t'^\ast)\,\dd t'^\ast+
\kappa^\ast\nabla^2_{\!z^\ast}\sigma^\ast
\end{eqnarray*}
which is exactly the expression given in
Eqs.~(\ref{spaceCHA})--(\ref{defRandM}) of the main text (dropping
stars out).
\section{Full analytical expression for mode equations}
\label{FullAnalytical}
In this Appendix, we give the full analytical expression of the
evolution equations for the stress modes $\sigma_n(t)$, of which
only a shortened version was given in \eq{ShortModeStressEq}. [For
memory modes, the analytical expression is straightforward and was
given in \eq{ShortModeMemoryEq}.]

The equation for a given mode $\sigma_n$ is found by projecting
the partial differential equation~(\ref{RheologicalEquation}) onto
the associated Fourier mode, i.e., by multiplying both sides by
$\cos(n\pi z/H)$ and integrating over $z$. Because the
non-linearities in $R(\sigma)$ are only polynomial, the integrals
are analytically tractable. (Note that this is a rare fact in
spectral or pseudo-spectral schemes; when non-linearities are more
complex, in general, exact expressions for mode equations do not
exist, and the numerical scheme accordingly increases in
complexity~\cite{Boyd}.) The only difficulty in the present case
comes from the fact that we deal with truncated Fourier series
(this explains some rather baroque expressions for summation
limits in the following).

Considering a truncation of order $N$ (i.e., the highest mode is
$\sigma_{N-1}$), the governing equation for the evolution of mode
$\sigma_n(t)$ writes (for all $n$ such that $0\leq n \leq N-1$):
\begin{widetext}
\begin{eqnarray*}
\dot\sigma_n&=& \delta^0_n\,\dot\gamma - a\,\sigma_n
+\frac{b}{2}\sum_{p=0}^n\sigma_p\sigma_{n-p} + (2 -
\delta_n^0)\frac{b}{2}\;\sum_{p=0}^{N-1-n}\sigma_p\sigma_{p+n}
-\frac{c}{4}\sum_{p=0}^n\sum_{q=0}^{n-p}\sigma_p\sigma_q\sigma_{n-p-q}
-\frac{c}{4}\sum_{p=0}^{N-1}\sum_{q=q_1}^{q_2}\sigma_p\sigma_q\sigma_{n+p-q}\\
& &-(3-2
\delta_n^0)\frac{c}{4}\;\sum_{p=0}^{N-1-n}~~\sum_{q=0}^{N-1-n-p}\sigma_p\sigma_q\sigma_{n+p+q}
-(2-\delta^0_n)\frac{c}{4}\sum_{p=0}^{N-1}\sum_{q=q_3}^{q_4}\sigma_p\sigma_q\sigma_{n+q-p}-\lambda
\,m_n - \kappa\Bigl(\frac{n\pi}{H}\Bigr)^2\sigma_n
\end{eqnarray*}
\end{widetext}
where $\delta$ is the usual Kronecker symbol, and the following
shorthands were used
\begin{eqnarray*}
q_1&=& \max\{n+p-N+1,0\}\\
q_2&=& N-1+\min\{n+p-N+1,0\}\\
q_3&=& \max\{0,p-n\}\\
q_4&=& N-1+\min\{0,p-n\}.
\end{eqnarray*}

In high-order truncations ($N=40$), a typical mode equation
contains several hundreds terms. Note also that the zeroth-mode
equation ($n=0$) is the only one where the shear rate $\dot\gamma$
appears; since, in addition, we have $\dot\sigma_0=\frac{\dd}{\dd
t}\avstress=0$ under conditions of fixed torque, this equation can
be used to compute $\dot\gamma(t)$ from the individual stress
modes $\sigma_n(t)$. This is in fact the same equation as
\eq{ShearRateEq}, but here the integral has been calculated explicitly.

\begin{thebibliography}{99}
%
\bibitem{Larson}
R.~G.~Larson, \emph{The Structure and Rheology of Complex Fluids}
(Oxford University Press, Oxford, 1999).
%
\bibitem{CatesHouches}
M.~E.~Cates, in \emph{Slow relaxations and nonequilibrium dynamics
in condensed matter, Les Houches Session LXXVII}, edited by
J.-L.~Barrat, M.~Feigelman, J.~Kurchan, and J.~Dalibard (Springer,
New York, 2003).
%
\bibitem{Spenley}
N.~A.~Spenley, M.~E.~Cates and T.~C.~B.~McLeish, \emph{Phys.\
Rev.\ Lett.} {\bf 71}, 939 (1993).
%
\bibitem{LuOlmstedBall}
C.-Y.~D.~Lu, P.~D.~Olmsted and R.~C.~Ball, \emph{Phys.\ Rev.\
Lett.} {\bf 84}, 642 (2000).
%
\bibitem{OlmstedEPL}
P.~D.~Olmsted, \emph{Europhys.\ Lett.} {\bf 48}, 339 (1999).
%
\bibitem{OlmstedGoveas}
J.~L.~Goveas and P.~D.~Olmsted, \emph{Eur.\ Phys.\ J.~E} {\bf 6},
79 (2001).
%
\bibitem{DrazinHydro}
P.~G.~Drazin and W.~H.~Reid, \emph{Hydrodynamic Stability}
(Cambridge University Press, Cambridge, 2004).
%
\bibitem{MannevilleLivre2}
P.~Manneville, \emph{Instabilities, Chaos and Turbulence}
(Imperial College Press, London, 2004).
%
\bibitem{ElasticTurbulence}
A.~Groisman and V.~Steinberg, \emph{Nature},  {\bf 405}, 53
(2000).
%
\bibitem{Hu}
Y.~T.~Hu, P.~Boltenhagen, E.~Matthys, and D.~J.~Pine,
\emph{J.~Rheol.} {\bf 42}, 1209 (1998).
%
\bibitem{Wheeler}
E.~K.~Wheeler, P.~Fischer, and G.~G.~Fuller, \emph{J.~Non-Newt.\
Fluid Mech.} {\bf 75}, 193 (1998).
%
\bibitem{Fischer1}
P.~Fischer, \emph{Rheol.~Acta} {\bf 39}, 234 (2000).
%
\bibitem{Fischer2}
P.~Fischer, E.~K.~Wheeler, and G.~G.~Fuller, \emph{Rheol.~Acta}
{\bf 41}, 35 (2002).
%
\bibitem{Fischer3}
V.~Herle, P.~Fischer, and E.~J.~Windhab, \emph{Langmuir} {\bf 21},
9051 (2005).
%
\bibitem{Wunenburger}
A.-S.~Wunenburger, A.~Colin, J.~Leng, A.~Arn{\'e}odo, and D.~Roux,
\emph{Phys.\ Rev.\ Lett.} {\bf 86}, 1374 (2001).
%
\bibitem{Salmon1}
J.-B.~Salmon, A.~Colin, and D.~Roux \emph{Phys.\ Rev.\ E} {\bf
66}, 031505 (2002).
%
\bibitem{Manneville}
S.~Manneville, J.-B.~Salmon, and A.~Colin, \emph{Eur.\ Phys.\
J.~E} {\bf 13}, 197 (2004).
%
\bibitem{Courbin}
L.~Courbin, P.~Panizza and J.-B.~Salmon, \emph{Phys.\ Rev.\ Lett.}
{\bf 92}, 018305 (2004).
%
\bibitem{Vlassopoulos}
L.~Hilliou and D.~Vlassopoulos, \emph{Ind.\ Eng.\ Chem.\ Res.}
{\bf 41}, 6246 (2002).
%
\bibitem{Laun}
H.~M.~Laun, \emph{J.\ Non-Newt.\ Fluid Mech.} {\bf 54}, 87 (1994).
%
\bibitem{Bandyopadhyay1}
R.~Bandyopadhyay, G.~Basappa, and A.~K.~Sood, \emph{Phys.\ Rev.\
Lett.} {\bf 84}, 2022 (2000).
%
\bibitem{Bandyopadhyay2}
R.~Bandyopadhyay and A.~K.~Sood, \emph{Europhys.\ Lett.} {\bf 56}
447 (2001).
%
\bibitem{Callaghan1}
W.~M.~Holmes, M.~R.~L{\'o}pez-Gonz{\'a}lez, and P.~T.~Callaghan,
\emph{Europhys.\ Lett.} {\bf 64}, 274 (2003).
%
\bibitem{Callaghan2}
M.~R.~L{\'o}pez-Gonz{\'a}lez, W.~M.~Holmes, P.~T.~Callaghan, and
P.~J.~Photinos, \emph{Phys.\ Rev.\ Lett.} {\bf 93}, 268302 (2004).
%
\bibitem{Salmon2}
J.-B.~Salmon, S.~Manneville, and A.~Colin, \emph{Phys.\ Rev.~E}
{\bf 68}, 051504 (2003).
%
\bibitem{Lootens}
D.~Lootens, H.~Van Damme, and P.~H{\'e}braud, \emph{Phys.\ Rev.\
Lett.} {\bf 90}, 178301 (2003).
%
\bibitem{CatesHead}
M.~E.~Cates, D.~A.~Head, and A.~Ajdari, \emph{Phys.\ Rev.~E} {\bf
66}, 025202 (2002).
%
\bibitem{HolmesJam}
C.~B.~Holmes, M.~E.~Cates, M.~Fuchs and P.~Sollich, \emph{J.\
Rheol.} {\bf 49}, 237 (2005).
%
\bibitem{Rings}
M.~E.~Cates and S.~J.~Candau, \emph{Europhys.\ Lett.} {\bf 55},
887 (2001).
%
\bibitem{AradianCatesEPL}
A.~Aradian and M.~E.~Cates, \emph{Europhys.\ Lett.} {\bf 70}, 397
(2005).
%
\bibitem{AradianProceedings}
A.~Aradian and M.~E.~Cates, in \emph{Proc. of the 3rd Int.\ Symp.\
on Slow Dynamics in Complex Systems}, edited by M.~Tokuyama and
I.~Oppenheim (American Institute of Physics, Melville, NY, 2004),
AIP Conference Proceedings vol. 708.
%
\bibitem{Fielding}
S.~M.~Fielding and P.~D.~Olmsted, \emph{Phys.\ Rev.\ Lett.} {\bf
92}, 084502 (2004).
%
\bibitem{Ramaswamy}
B.~Chakrabarti, M.~Das, C.~Dasgupta, S.~Ramaswamy, and A.~K.~Sood,
\emph{Phys.\ Rev.\ Lett.} {\bf 92}, 055501 (2004).
%
\bibitem{StAndrews}
\emph{Soft and Fragile Matter: Nonequilibrium Dynamics, Metastability
and Flow}, edited by M.~R.~Evans and M.~E.~Cates (Insitute of
Physics Publishing, Bristol, 2000).
%
\bibitem{SGRPRL}
P.~Sollich, F.~Lequeux, P.~H{\'e}braud, and M.~E.~Cates {\bf 78},
2020 (1997).
%
\bibitem{HeadSGR}
D.~A.~Head, M.~E.~Cates, and A.~Ajdari, \emph{Phys.\ Rev.~E} {\bf
64}, 061509 (2001).
%
\bibitem{Fluidity}
C.~Derec, A.~Ajdari and F.~Lequeux, \emph{Eur.\ Phys.~J.~E} {\bf
4}, 355 (2001).
%
\bibitem{FieldingPlanarInterface}
S.~M.~Fielding, \emph{Phys.\ Rev.\ Lett.} {\bf 95}, 134501 (2005).
%
\bibitem{ChaikinLubensky}
P.~M.~Chaikin and T.~C.~Lubensky, \emph{Principles of condensed
matter physics} (Cambridge University Press, Cambridge, 1995).
%
\bibitem{FitzHugh}
R.~FitzHugh, \emph{Biophys.~J.} {\bf 1}, 445 (1961).
%
\bibitem{Nagumo}
J.~Nagumo, S.~Yoshizawa, and S.~Arimoto, \emph{Proc.\ IRE } {\bf
50}, 2061 (1965).
%
\bibitem{Murray}
J.~D.~Murray, \emph{Mathematical Biology}, 3rd Ed. (Springer
Verlag, New York, 2002)
%
\bibitem{KeenerSneyd}
J.~Keener and J.~Sneyd, \emph{Mathematical Physiology} (Springer
Verlag, New York, 1998).
%
\bibitem{Rocsoreanu}
C.~Rocsoreanu, A~Georgescu, and N.~Giurgiteanu, \emph{The
FitzHugh-Nagumo model} (Kluwer Academic Publishers, Amsterdam,
2000).
%
\bibitem{FHN1}
J.~A.~Acebr{\'o}n, A.~R.~Bulsara, and W.-J.~Rappel,  \emph{Phys.\
Rev.~E} {\bf 69}, 026202 (2004).
%
\bibitem{FHN2}
G.~Baier and M.~M{\"u}ller, \emph{Rev.\ Mex. F{\'i}sica} {\bf 50}
422 (2004).
%
\bibitem{FHN3}
C.~G.~Assisi, V.~K.~Jirsa, and J.~A.~S.~Kelso,
\emph{Phys.\ Rev.\ Lett.} {\bf 94}, 018106 (2005).
%
\bibitem{FHN4}
M.~A.~Zaks, X.~Sailer, L.~Schimansky-Geier, and A.~B.~Neiman,
\emph{Chaos} {\bf 15}, 026117 (2005).
%
\bibitem{FieldingConcentration}
S.~M.~Fielding and P.~D.~Olmsted, \emph{Eur.\ Phys.~J.~E} {\bf
11}, 65 (2003).
%
\bibitem{Boyd}
J.~P.~Boyd, \emph{Chebyshev and Fourier Spectral Methods} (Dover
Publications, New York, 2000).
%
\bibitem{ElKareh}
A.~W.~El-Kareh and L.~G.~Leal, \emph{J.~Non-Newton.\ Fluid Mech.}
{\bf 33}, 257 (1989).
%
\bibitem{Mathematica}
\emph{Mathematica}, Version 5.0 (Wolfram Research, Inc.,
Champaign, 2003).
%
\bibitem{Glendinning}
P.~Glendinning, \emph{Stability, Instability and Chaos} (Cambridge
University Press, Cambridge, 1994).
%
\bibitem{ChemicalOscillators}
I.~R.~Epstein and J.~A.~Pojman, \emph{An Introduction to Nonlinear
Chemical Dynamics} (Oxford University Press, Oxford, 1998).
%
\bibitem{MannevilleLivre1}
P.~Manneville, \emph{Dissipative Structures and Weak Turbulence}
(Academic Press, New York, 1990).
%
\bibitem{AjdariInterface}
A.~Ajdari, \emph{Phys.\ Rev.~E} {\bf 58}, 6294 (1998).
%
\end{thebibliography}
\end{document}